%% file: template.tex
\documentclass{article}

\usepackage{arxiv}

\usepackage[utf8]{inputenc} % allow utf-8 input
\usepackage[T1]{fontenc}    % use 8-bit T1 fonts
\usepackage{hyperref}       % hyperlinks
\usepackage{url}            % simple URL typesetting
\usepackage{booktabs}       % professional-quality tables
\usepackage{amsfonts}       % blackboard math symbols
\usepackage{nicefrac}       % compact symbols for 1/2, etc.
\usepackage{microtype}      % microtypography
\usepackage{lipsum}		% Can be removed after putting your text content
\usepackage{natbib}
\newtheorem{scenario}{Scenario}
\usepackage{doi}
\input{lib/preambule.tex}

\title{Multi-self-loop Lackadaisical Quantum Walk with Partial Phase Inversion}

\author{ \href{https://orcid.org/0000-0002-6527-7065}{\includegraphics[scale=0.06]{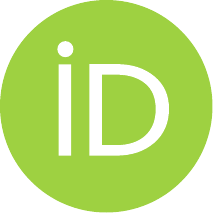}\hspace{1mm}Luciano S. de Souza}\thanks{R. Dom Manuel de Medeiros, s/n, Dois Irmãos -- Recife, Pernambuco -- Brasil} \\
	Departamento de Estat\'{i}stica e Inform\'{a}tica\\
	Universidade Federal Rural de Pernambuco\\
	Recife, Brasil \\
	\texttt{luciano.serafim@ufrpe.br} \\
	\And
	\href{https://orcid.org/0000-0002-2672-7801}{\includegraphics[scale=0.06]{orcid.pdf}\hspace{1mm}Jonathan H. A. de Carvalho} \\
	Centro de Inform\'{a}tica\\
	Universidade Federal de Pernambuco\\
	Recife, Brasil \\
	\texttt{jhac@cin.ufpe.br} \\
 \And
	\href{https://orcid.org/0000-0002-9544-7774}{\includegraphics[scale=0.06]{orcid.pdf}\hspace{1mm}Henrique C. T. Santos} \\
 Instituto Federal de Educação, Ciência e\\ Tecnologia de Pernambuco\\
	Recife, Brasil \\
	\texttt{henrique.santos@recife.ifpe.edu.br}
	\And
	\href{https://orcid.org/0000-0002-2131-9825}{\includegraphics[scale=0.06]{orcid.pdf}\hspace{1mm}Tiago A. E. Ferreira} \\
	Departamento de Estat\'{i}stica e Inform\'{a}tica\\
	Universidade Federal Rural de Pernambuco\\
	Recife, Brasil \\
	\texttt{tiago.espinola@ufrpe.br} \\
}

\renewcommand{\shorttitle}{MSLQW with Partial Phase Inversion}

\hypersetup{
pdftitle={A template for the arxiv style},
pdfsubject={q-bio.NC, q-bio.QM},
pdfauthor={David S.~Hippocampus, Elias D.~Striatum},
pdfkeywords={First keyword, Second keyword, More},
}

\begin{document}
\maketitle

\begin{abstract}
The lackadaisical quantum walk, a quantum analog of the lazy random walk, is obtained by adding a weighted self-loop transition to each state. Impacts of the self-loop weight $l$ on the final success probability in finding a solution make it a key parameter for the search process. The number of self-loops can also be critical for search tasks. This article proposes the quantum search algorithm Multi-self-loop Lackadaisical Quantum Walk with Partial Phase Inversion, which can be defined as a lackadaisical quantum walk with multiple self-loops, where the target state phase is partially inverted. In the proposed algorithm, each vertex has $m$ self-loops, with weights $l' = l/m$, where $l$ is a real parameter. The phase inversion is based on Grover's algorithm and acts partially, modifying the phase of a given quantity $s < m$ of self-loops. On a hypercube structure, we analyzed the situation where $1 \leqslant m \leqslant 30$. We also propose two new weight values based on two ideal weights $l$ used in the literature. We investigated the effects of partial phase inversion in the search for $1$ to $12$ marked vertices. As a result, this proposal improved the maximum success probabilities to values close to $1$ in $O (\sqrt{(n+m)\cdot N})$, where $n$ is the hypercube degree. This article contributes with a new perspective on the use of quantum interferences in constructing new quantum search algorithms.
\end{abstract}

% keywords can be removed
\keywords{Quantum Computing \and Quantum Interference \and Quantum Walks \and Quantum Search Algorithm \and Lackadaisical Quantum Walk \and Multiple Self-loops \and Partial Phase Inversion}

\section{Introduction}
\label{sec:introduction}

Just as quantum interference plays an essential role in the development of quantum algorithms, for \citet{shenvi2003quantum}, quantum walks provide one of the most promising features, an intuitive framework for building new quantum algorithms. Their pioneering paper designed a quantum search algorithm based on quantum walks. They demonstrated that this quantum search algorithm could be used to find a marked vertex of a hypercube. Although they are distinct algorithms, there are several similarities with Grover's algorithm~\citep{grover1996fast}. Both algorithms start in the state of equal superposition over all states. They make use of Grover's diffusion operator. They can be seen as a rotation in a two-dimensional subspace. They have the same running time. The measurement is performed at a specific time to obtain the maximum probability of success. Both algorithms use an oracle that marks the target state with a phase of $-1$, \textit{i.e.}, they use quantum interference to develop their quantum search algorithms. 

Since then, some works have been developed to improve the capacity of the quantum search algorithm in the hypercube \citep{potovcek2009optimized, hein2009quantum}. The addition of self-loops at each vertex is part of many improvement proposals for quantum search algorithms. \citet{hoyer2009faster} used a directed walk on the line to produce an algorithm with $(N - 1)$ self-loops at each vertex, where $N$ is the total number of vertices. In its turn, \citet{wong2015grover} proposed a quantum analog of the classical lazy random walk, called the lackadaisical quantum walk - LQW. The quantum walker has a chance to stay at the same vertex by introducing $m$ integer self-loops at each vertex of the graph, and its effects were investigated with Grover's search algorithm. However, \citet{wong2017coined} proposed a modification that reduced the number $m$ to a single self-loop with a non-integer weight $l$.

Recently, some works have investigated the application of the lackadaisical quantum walk on the hypercube structure. Rhodes and \citet{rhodes2020search} showed that the ideal value for the self-loop weight in the search for a single marked vertex is $l = d/N$, where $d$ is the degree of the vertex and $N$ is the number of nodes in the hypercube. \citet{souza2021lackadaisical} showed that the optimal self-loop weight value for searching a single marked vertex is not optimal for searching multiple marked vertices on the hypercube. Thus, they defined a new value of $l$ which is the value proposed by Rhodes and \citet{rhodes2020search} multiplied by the number of marked vertices $k$, resulting in $l = (d/N)\cdot k$.

Although \citet{souza2021lackadaisical} have obtained improvements in the search for multiple vertices in the hypercube by proposing a new ideal weight value for the self-loop, in some cases, it is not possible to maintain the performance of the quantum search algorithm using the LQW. The LQW's original proposal used $m$ self-loops with an integer weight $l$, and later this number was reduced to a single self-loop with a non-integer weight. However, according to Rhodes and \citet{rhodes2020search}, if the weight value $l$ of the self-loop is an integer, it is equivalent to having several unweighted self-loops. We can therefore suggest that there is a relationship between the weight value $l$ and the number $m$ of self-loops.

Therefore, in the described scenario, it is possible to observe three aspects that were discussed, which can interfere with the performance of the LQW. The first is quantum interference, which in practical terms, can be achieved with the phase inversion of a target state. The second is the weight value of the self-loop, and the third is the number of self-loops. Another aspect we must consider is the distribution of the weight value between the multiple self-loops. Some authors use different strategies to define how the weights are distributed in the set of vertices.

In the work developed by \citet{hoyer2009faster}, the amplitude of each self-loop is the same, and they are grouped into a single normalized state. \citet{wang2017adjustable} proposed a model of adjustable self-loops controlled by a real parameter in the coin operator. Rhodes and \citet{rhodes2019quantum} investigated the spatial search in the complete bipartite graph, which can be irregular with $N_{1}$ and $N_{2}$ vertices in each partition. In this way, self-loops in each set of vertices can have different weights $l_{1}$ and $l_{2}$, respectively. Rapoza and \citet{rapoza2021search} defined a self-loop weight value for marked vertices, while the remaining weights for unmarked vertices can be chosen randomly.

Based on the previous information and through experiments, we verified that using a single self-loop at each hypercube vertex with a non-integer weight $l$ is equivalent to using $m$ self-loops where each self-loop has a non-integer weight $l/m$. Considering Grover's search algorithm used in the LQW, in practical terms, after executing the part of the quantum system responsible for detecting the target state, a phase rotation is applied, \textit{i.e.}, by linearity, it inverts the phase of all the base states related to the target state and applies the diffusion transformation.

According to \citet{mcmahon2007quantum}, quantum interference plays an important role in the development of quantum algorithms. There are two types of interference, positive (constructive) and negative (destructive), in which the probability amplitudes add constructively or the probability amplitudes add destructively, respectively. For \citet{grover1996fast}, which developed an algorithm based on quantum interference, the operation that results in the phase shift of the target state is one of the procedures that form the basis of quantum mechanics algorithms and makes them more efficient than classical analog algorithms.

However, as seen previously, in some cases, it is impossible to maintain the performance of the LQW. Since the $m$ self-loops are redundant, the proposal of this work is a different way of performing the phase inversion process. We suggest modifying Grover's search algorithm to make possible the partial inversion of the base states that represent the $m$ self-loops of the marked vertices. Therefore, we propose the Multi-self-loop Lackadaisical Quantum Walk with Partial Phase Inversion - MSLQW-PPI. We will show and analyze the effects of using multiple real-valued weighted self-loops at each vertex of an n-dimensional hypercube with Grover's search algorithm using a different strategy in the phase inversion operation that only acts on $s < m$ self-loops.

The LQW algorithm highly depends on the self-loop weight value. The ideal weight composition in several structures, including the hypercube, considers the vertex degree, the total number of vertices, and the number of marked vertices~\citep{wong2018faster,rhodes2019search,giri2020lackadaisical,de2023applying}. Therefore, based on the weights proposed by Rhodes and \citet{rhodes2020search} and \citet{souza2021lackadaisical} for the use of a single self-loop, we also suggest two new weights for the use of multiple self-loops that explore this relationship between vertex degree, the total number of vertices and the number of marked vertices. We added an integer exponent $\alpha$ equal to $2$, restricting the analysis to this value as the initial choice. In this way, we can analyze the performance of the quantum walk in the hypercube, maintaining the ideal weights' composition, just by modifying its scale.

The main contributions of this work are summarized as follows. We present a new quantum search algorithm based on lackadaisical quantum walks. For this, we revisit the use of multiple self-loops per vertex and propose a partial phase inversion of the target states based on a modification in Grover's oracle. We formulate two new weight values $l$, such that $l = n^{2}/N$ and $l = (n^{2}/N)\cdot k$, where $n^{2}$ is the degree of the vertex in a n-regular structure $2$ times, $N$ is the number of vertices, and $k$ is the number of marked vertices. Each $m$ self-loop is weighted in the form $l' = l/m$. Finally, with this approach, we were able to increase the maximum success probabilities to values close to $1$.

This paper is organized as follows. In Section \ref{sec:quantum-walks-hypercube}, we present some concepts about quantum walks in the hypercube. In Section \ref{sec:article-proposal}, we present the proposal for this work. In Section \ref{sec:experiments-setup}, the experiments are defined. Section \ref{sec:results-and-discussion} presents the results and discussion. Finally, Section \ref{sec:conclusions} contains the conclusions.

\section{Lackadaisical quantum walk on the hypercube}
\label{sec:quantum-walks-hypercube}

Quantum walks are the quantum counterpart of classical random walks. They are an advanced tool that provides one of the most promising features, an intuitive framework for building new quantum algorithms \citep{aharonov1993quantum, shenvi2003quantum, ambainis2012search}. The evolution of the discrete-time quantum walk occurs by successive applications of a unitary evolution operator $U$ that operates in the Hilbert space,

\begin{align}
\label{eq:hilbert-space-quantum-walks}
\mathcal{H} = \mathcal{H}^{C} \otimes \mathcal{H}^{S}
\end{align}
where the coin space $\mathcal{H}^{C}$ is the Hilbert space associated with a quantum coin, and the walker space $\mathcal{H}^{S}$ is the Hilbert space associated with its position representation. The evolution operator $U$ is defined as follows,

\begin{align}
\label{eq:evolution-operator}
    U = S(C\otimes I_{N})
\end{align}
where $S$ is the shift operator that acts in the walker's space based on the state of the coin. $I_{N}$ is the identity matrix, and the unitary matrix $C$ is the coin operator \citep{shenvi2003quantum}. The evolution equation represented by a quantum walk at time $t$ is given by

\begin{align}
\ket{\Psi(t)} = U^{t}\ket{\Psi(t=0)}.
\end{align}

In general, the study of quantum walks needs a structure to represent the time evolution of the walker. It is possible to use many different structures, such as complete and johnson's graphs \citep{wong2015grover,wong2017coined,ZHANG20181}, grids \citep{saha2022faster,de2023applying}, hypercubes \citep{rhodes2020search,souza2021lackadaisical}, among others \citep{rhodes2019quantum,tanaka2022spatial,qu2022deterministic}. Here, it was chosen the hypercube structure. The hypercube was used by \citet{shenvi2003quantum} as a temporal evolution structure to propose in his pioneering work a search algorithm based on quantum walks. Furthermore, a quantum walk on the hypercube can be reduced to a quantum walk on a line which reduces its complexity. The n-degree hypercube is an undirected graph with $2^{n}$ nodes, where each node can be described by a binary string of $n$ bits. In this way, two nodes $\Vec{x}$ and $\Vec{y}$ in the hypercube are adjacent if $\Vec{x}$ and $\Vec{y}$ differ by only a single bit, \textit{i.e.}, if $|x - y| = 1$, where $|x - y|$ is the Hamming distance between $\Vec{x}$ and $\Vec{y}$ \citep{venegas2012quantum}.

Let us define the Hilbert space associated with the quantum walk on the hypercube. According to Equation \ref{eq:hilbert-space-quantum-walks}, the Hilbert space associated with the quantum coin space is $\mathcal{H}^{C}$, and the Hilbert space associated with the walker's position is $\mathcal{H}^{S}$. Then, the Hilbert space associated with the quantum walk in the hypercube is

\begin{align}
    \mathcal{H} = \mathcal{H}^{n} \otimes \mathcal{H}^{2^{n}}
\end{align}
where $\mathcal{H}^{n}$ is the Hilbert space associated with the quantum coin space, and $\mathcal{H}^{2^{n}}$ is the Hilbert space associated with nodes in the hypercube, which represents the walker's position. In an n-dimensional hypercube, the $i$ directions define the states of the coin and can be labeled by the $n$ base vectors $\{\ket{0}, \ket{1}, \dots, \ket{n-1} \}$. Each one of these $n$ base vectors can be represented by $\{\ket{e_{0}},\ket{e_{1}},\dots,\ket{e_{n-1}}\}$, where $e_{i}$ is a binary string of $n$ bits with $1$ in the $i$-th position \citep{kempe2002quantum,shenvi2003quantum}. The shift operator $S$, described in Equation \ref{eq:shift-operator-hypercube}, acts mapping a state $\ket{i,\vec{x}} \rightarrow \ket{i,\vec{x}\oplus \vec{e_{i}}}$.

\begin{align}
\label{eq:shift-operator-hypercube}
    S = \sum_{i=0}^{n-1}\sum_{\vec{x}} \ket{i,\vec{x}\oplus \vec{e_{i}}}\bra{i,\vec{x}}
\end{align}

The pure quantum walk (without search) evolves by repeated applications of the evolution operator described in Equation \ref{eq:evolution-operator}, where $C$ is Grover's ``diffusion'' operator on the coin space, and it is given by

\begin{align}
\label{eq:grovers-coin}
    C = 2\ket{s^{C}}\bra{s^{C}} - I_{n}
\end{align}
where $I_{n}$ is the identity operator, and $\ket{s^{C}}$ is the equal superposition over all $n$ directions \citep{moore2002quantum,shenvi2003quantum}, \textit{i.e.},

\begin{align}
    \ket{s^{C}} = \frac{1}{\sqrt{n}} \sum_{i=0}^{n-1} \ket{i}.
\end{align}

Now, consider the quantum walk with search. A query to the ``Grover oracle'', described in Equation \ref{eq:query-oracle-grover}, is included in each step of the quantum walk.

\begin{align}
\label{eq:query-oracle-grover}
    U' = U \cdot (I_{n} \otimes Q)
\end{align}
where $Q = I_{N} - 2\ket{\omega}\bra{\omega}$, and $\ket{\omega}$ is the marked vertex. The initial state of the quantum walk in the hypercube is defined according to Equation \ref{eq:initial-state-quantum-walk-hypercube} as an equal superposition for all $N$ nodes and $n$ directions.

\begin{align}
\label{eq:initial-state-quantum-walk-hypercube}
    \ket{\Psi(t = 0)} = \frac{1}{\sqrt{n}} \sum_{i=0}^{n-1}\ket{i}\otimes \frac{1}{\sqrt{N}} \sum_{\vec{x}} \ket{\vec{x}}
\end{align}

Finally, we will define the lackadaisical quantum walk in the hypercube. The lackadaisical quantum walk is a quantum analog of the lazy random walk. This quantum algorithm is obtained by adding at least one self-loop to each graph vertex \citep{wong2015grover}. According to the definition presented by \citet{hoyer2020analysis}, considering an n-regular graph with a single marked vertex, by adding a self-loop of weight $l$ to each vertex, the coined Hilbert space becomes

\[\mathcal{H}^{n+1} = \{\ket{e_{0}}, \ket{e_{1}}, \dots, \ket{e_{n-1}}, \ket{\circlearrowleft}\},\]
where $\ket{\circlearrowleft}$ represents the self-loop. Thus weighted self-loop accounting is done by modifying Grover's coin presented in Equation~\ref{eq:grovers-coin},  as follows

\begin{align}
\label{eq:grovers-coin-self-loop}
    C = 2\ket{s^{C}}\bra{s^{C}} - I_{(n + 1)}
\end{align}
where

\begin{align}
\label{eq:s-c-add-self-loop}
    \ket{s^{C}} = \frac{1}{\sqrt{n + l}} \left ( \sqrt{l}\ket{\circlearrowleft} + \sum_{i=0}^{n-1}\ket{i} \right ).
\end{align}

\section{Article proposal}
\label{sec:article-proposal}

Let us present the proposal for this work, the Multi-self-loop Lackadaisical Quantum Walk with Partial Phase Inversion (MSLQW-PPI). It is an alternative to exploring multiple non-integers self-loops in a way that can improve the results of the lackadaisical quantum walk. Considering an n-regular hypercube, we add $m$ self-loops with weights $l'$ at each vertex, \textit{i.e.}, the amplitude of each self-loop is weighted by $\sqrt{l'}$. Therefore, the Hilbert space associated with the coin space becomes

\begin{align*}
    \mathcal{H}^{n+m} &= \{\ket{e_{0}}, \ket{e_{1}}, \dots, \ket{e_{n-1}},  \ket{\circlearrowleft_{0}}, \ket{\circlearrowleft_{1}}, \dots, \ket{\circlearrowleft_{\text{m}-1}}\}.
\end{align*}

To account for the $m$ weighted self-loops, a new modi\-fication was made to Grover's coin described in Equation~\ref{eq:grovers-coin-self-loop}, as follows,

\begin{align}
\label{eq:grovers-coin-m-self-loop}
    C = 2\ket{s^{C}}\bra{s^{C}} - I_{(n + m)}
\end{align}
where

\begin{align}
\label{eq:add-mult-self-loop}
    \ket{s^{C}} = \frac{1}{\sqrt{n + l}} \left (\sqrt{l'} \sum_{j=0}^{m-1} \ket{\circlearrowleft_{j}} + \sum_{i=0}^{n-1}\ket{i} \right )
\end{align}
and $l' = l/ m$. In this way, according to Equation \ref{eq:add-mult-self-loop}, we consider that the weight $l$ has its value equally distributed to the $m$ self-loops. The MSLQW-PPI system in the hypercube starts as follows,

\begin{align}
\label{eq:initial-system-mslqw}
    \ket{\Psi(t = 0)} = \ket{s^{C}} \otimes \frac{1}{\sqrt{N}} \sum_{\vec{x}}\ket{\vec{x}}.
\end{align}

Substituting Equation \ref{eq:add-mult-self-loop} into Equation \ref{eq:initial-system-mslqw} and applying the expansions, we obtain the initial state described in Equation \ref{eq:initial-state-mslqw}.

\begin{align}
\label{eq:initial-state-mslqw}
\begin{split}
\ket{\Psi(t = 0)} &= \frac{\sqrt{l'}}{\sqrt{n+l} \times \sqrt{N}} \sum_{j=0}^{m-1}\sum_{\vec{x}} \ket{\circlearrowleft_{j},\vec{x}} + \frac{1}{\sqrt{n+l} \times \sqrt{N}} \sum_{i=0}^{n-1}\sum_{\vec{x}}\ket{i,\vec{x}}
\end{split}
\end{align}

To properly explore those multiple self-loops at each vertex, we propose a modification to Grover's oracle described previously.
Note that, in Equation \ref{eq:query-oracle-grover}, a query $(I_{n}\otimes Q)$ is included at each step of the quantum walk, where $Q = I_{N} - 2 \ket{\omega}\bra{\omega}$ and $\omega$ is the marked vertex. Thus, by linearity, when we apply the oracle to all $\ket{i}\ket{\Vec{x}}$ states of the superposition of vertices and edges, there are two possibilities. The first possibility is that $\Vec{x}$ is not the target state, \textit{i.e.}, $\omega \neq \Vec{x}$. As $\braket*{\omega}{\Vec{x}} = 0$, the state stays unchanged. The second possibility is that $\Vec{x}$ is the target state, \textit{i.e.}, $\omega = \Vec{x}$. In this case, as $\braket*{\omega}{\Vec{x}} = 1$, we have the phase inversion of the target state.

As can be seen, this oracle depends exclusively on the vertex in question. The edge is not considered for state phase inversion. All states associated with the vertex in question, independent of the edge, are also inverted. Here, we propose a partial inversion of the states related to the marked vertices. According to \citet{rhodes2020search}, the number of self-loops is a parameter that adjusts the probability of a walker staying put. The idea here is to be able to invert the phase of a target self-loops $\circlearrowleft_{\tau}$ and all edges that are not self-loops $\epsilon$ of the hypercube of a target vertex and investigate their effects. Inspired by \citet{hoyer2009faster}, we identify each edge of the hypercube by assigning a basis vector, as follows, $\ket{\circlearrowleft_{j}, \vec{x}}$ and $\ket{i, \Vec{x}}$, where $0 \leqslant j \leqslant m-1$ and $0 \leqslant i \leqslant n-1$. Hence, each state $\ket{\textbf{x}}$, which represents a walker's position, is a linear combination of the states,

\begin{align}
\begin{split}
\label{eq:arbitrary-state}
\ket{\textbf{x}} &= \ket{\circlearrowleft_{0},\vec{x}}+ \dots + \ket{\circlearrowleft_{\text{m}-1},\vec{x}}+ \ket{0,\vec{x}}+\dots+ \ket{n-1,\vec{x}}
\end{split}
\end{align}
which denotes the superposition of all edges \citep{yu2018searching}. In the case where $\ket{\textbf{x}}$ contains the target state, it will have the phase of the components $\ket{\circlearrowleft_{\tau},\Vec{x}}$ and $\ket{\epsilon,\Vec{x}}$ changed. It requires an oracle that identifies the state's components. Consider, again, Grover's oracle described in Equation \ref{eq:query-oracle-grover}. The proposed modification of the oracle described in Equation \ref{eq:oracle-mslqw} makes it possible to identify the components of the target state.

\begin{align}
\begin{split}
\label{eq:oracle-mslqw}
   Q &= I_{(n+\text{m})\cdot N}  - 2 \sum_{\omega}\sum_{\epsilon = 0}^{n-1} \ket{\epsilon,\omega}\bra{\epsilon,\omega}- 2\sum_{\omega}\sum_{\tau} \ket{\circlearrowleft _{\tau},\omega}\bra{\circlearrowleft _{\tau},\omega} 
\end{split}
\end{align}
where $\ket{\omega}$ represents the marked vertex, $\epsilon$ represents an edge that is not a self-loop, and $\circlearrowleft_{\tau}$ are the self-loops that will have their phases inverted. In this way, $Q$ acts in the coin and vertex space described in Equation \ref{eq:hilbert-space-quantum-walks}, contrary to Equation \ref{eq:query-oracle-grover}, where $Q$ acts in the vertex space. Consider an arbitrary state $\ket{\textbf{x}}$. When applying the proposed oracle, the phase of the components that represent their edges is considered individually. Considering Equation \ref{eq:modified-oracle-application}, at each step of the quantum walk an oracle query is applied to each edge of the state by linearity.

\begin{align}
\label{eq:modified-oracle-application}
\begin{split}
       Q \ket{\textbf{x}} &=  Q\ket{\circlearrowleft_{0},\Vec{x}} + Q\ket{\circlearrowleft_{1},\Vec{x}} + \cdots + Q\ket{\circlearrowleft_{\text{m}-1},\Vec{x}}+ Q\ket{0,\Vec{x}}+Q\ket{1,\Vec{x}}+\dots+Q\ket{n-1,\Vec{x}}
\end{split}
\end{align}
Here also there are two possibilities. The first possibility is that $\ket{\textbf{x}}$ does not contain the target state. The second possibility is that $\ket{\textbf{x}}$ contains the target state. Finally, it is necessary to define the self-loop to have its phase inverted.
Here, the phase of $s$ self-loops is inverted, where experiments with $1 \leqslant s \leqslant m$ were done. The simpler situation where only one single self-loop is inverted can be used to illustrate the process. Consider the states $\ket{\circlearrowleft_{\tau=j}}$ as the target self-loops without loss of generality. Therefore, the description of the oracle is done as follows,

\begin{align}
\begin{split}
\label{eq:oracle-mslqw-self-loop-defined}
   Q &= I_{(n+\text{m})\cdot N} - 2\sum_{\omega}\sum_{\epsilon = 0}^{n-1}\ket{\epsilon,\omega}\bra{\epsilon,\omega}-2\sum_{\omega}\sum_{\tau}\ket{\circlearrowleft_{\tau=j},\omega}\bra{\circlearrowleft_{\tau=j},\omega}
\end{split}
\end{align}
For the case where $\ket{\textbf{x}}$ does not contain the target state, the phase of the states remains unchanged. For the case where $\ket{\textbf{x}}$ contains the target state, we have the partial phase inversion of the components $\ket{\circlearrowleft_{\tau=j},\Vec{x}}$ and $\ket{i,\Vec{x}}$, while the components $\ket{\circlearrowleft_{\tau\neq j},\Vec{x}}$ stays unchanged. Additional details on the application of the new oracle in three possible scenarios can be found in Appendix~\ref{sec:appendix-oracle-evaluation}.

\section{Experiment setup}
\label{sec:experiments-setup}

According to the definitions of the hypercube, two marked vertices are adjacent if the Hamming distance between them is $1$. A set of non-adjacent marked vertices have a Hamming distance of at least $2$ from any other marked vertex, \textit{i.e.}, they are mutually non-adjacent. In the experiments performed in this paper, we consider only the scenario where the marked vertices are non-adjacent. The search for adjacent marked vertices constitutes a separate scenario and will not be addressed in this work.

Experiments were performed to analyze the behavior of a Lackadaisical Quantum Walk with multiple self-loops at each vertex, where, the quantity $s$ $(1 \leqslant s \leqslant m)$ of self-loops is inverted. The case $s = m$ is the conventional oracle operation situation. Initially, we used the weight values proposed by \citet{rhodes2020search} and \citet{souza2021lackadaisical}. These experiments were also made for the weight values proposed in this work. To evaluate the relative dispersion behavior of the mean success probability, we used Pearson's coefficient of variation (the ratio between the standard deviation and the mean value).

\subsection{Definition of vertex sets and simulations}
\label{sec:definition-of-vertex-sets-and-simulations}

To determine how the simulations are performed, it is necessary to define how the marked vertices are divided. For each number of marked vertices $k$, $\gamma$ simulations are performed varying the position of the $k$ marked vertices. Therefore, the marked vertices are divided into groups of $M_{k,\gamma}$ samples. Here we define $1 \leqslant k \leqslant 12$ and $\gamma = 100$. In this way, we have a set of twelve hundred samples. This set is divided into twelve groups of one hundred samples as follows: $M_{1,100}, M_{2,100}, M_{3,100}, \cdots, M_{12,100}$. Each sample was made without replacement following a uniform distribution, \textit{i.e.}, each one of them has $k$ distinct vertices. For each group of one hundred samples, we fixed the $k$ number of marked vertices and vary their location, for example, when $k = 2$, as shown below,

\begin{align*}
    M_{2,100} &= [\{254, 1498\}_{1},\{969, 3520\}_{2}, \dots,\{410, 1121\}_{100}].
\end{align*}
The values shown, for example, $\{254, 1498\}_{1}$, are the 1st sample from a total of $100$ with $2$ marked vertices, where its binary representation is $\{000011111110, 010111011010\}_{1}$ in the hypercube. In this way, for every new group of simulations $M_{k,\gamma}$, $k\cdot100$ non-adjacent vertices are marked and divided into $100$ samples of $k$ vertices. Each one of these samples with $k$ vertices is a computational experiment. For instance, $ k = 3$
\begin{align*}
    M_{3,100} &= [\{3034, 1616, 2438\}_{1},\{2059, 2745, 3686\}_{2},\dots,\{3017, 3484, 1773\}_{ 100}].
\end{align*}
As a first investigation, the experiments employ $s=1$ inverted self-loop. Experiments were also performed for full-phase inversion where $s=m$. In the second investigation, the experiments employ $2 \leqslant s < m $. Therefore, $1 \leqslant s \leqslant 30$ and $1 \leqslant m \leqslant 30$. For each sample $M_{k,\gamma}$, all possible combinations of $s \leqslant m$ are done. For example, when $m=1$, then $s=1$. When $m=2$, then $s=1$, and $s=2$, and so on until $m=30$, the $s = \{1,2,\cdot,30\}$. This way,  a total of $558000$ simulations were performed. A simulation stops after each quantum walk has reached the number of iterations necessary to obtain the maximum value of the probability amplitude in the $k$ marked vertices. Considering $s=1$ and $m=1$ we have the same conditions as \citet{rhodes2020search} and \citet{souza2021lackadaisical}, \textit{i.e.} the use of a single self-loop.

\subsection{Hardware and software setup}
\label{sec:hardware-software-setup}

The simulations were performed using the Parallel Experiment for Sequential Code - PESC~\citep{santos2023pesc}. The PESC is a computational platform for distributing computer simulations on the resources available on a network by packaging the user code in containers that abstract all the complexity needed to configure these execution environments, allowing any user to benefit from this infrastructure. All client nodes that participated in the simulations use the Ubuntu 18.04.6 LTS (Bionic Beaver) operational system and have an HD of 500 GB. Other machine settings are shown in Table~\ref{tab:machine-settings}.

\begin{table*}[]
\centering
\caption{Client machine settings.}
\begin{tabular}{lrl}
\toprule
\textbf{Node} & \multicolumn{1}{l}{\textbf{System RAM}} & \textbf{System Processor}                           \\ \midrule
Node 1 and 2  & 32 GB                                  & Intel(R) Core(TM) i7-2600K CPU @ 3.40GHz            \\
Node 3        & 8 GB                                 & Intel(R) Core(TM) i7-2600 CPU @ 3.40GHz             \\
Node 4 and 5  & 32 GB                                  & Intel(R) Core(TM) i7-6700 CPU @ 3.40GHz             \\
Node 6        & 16 GB                                  & Intel(R) Core(TM) i7-8700 CPU @ 3.20GHz \\ \bottomrule
\end{tabular}
\label{tab:machine-settings}
\end{table*}

The PESC platform provides a web interface where the user configures a request to execute a simulation. In this request, the user can inform the number of times the simulation must be repeated. Each instance of the simulation receives via parameter the instance identification, called a rank. The rank is used to initialize variables and parameterize other processes. The programming language used to write the algorithms was Python 3.7.

After receiving the request, the PESC platform distributes the simulations to the available client nodes. The distribution is based on each client's workload, where each client's performance factor is informed when connecting to the server. Before starting the execution, each client creates and configures the environment necessary to run the received code. The user informs the execution environment needs at the request moment.

Using the platform simplified the simulation execution process as it manages the status and life cycle of the request, restarting simulation instances in case of failure on any of the clients, with the possibility of moving the instance to another client node depending on the type of failure detected.

All this was very important considering the nature of the proposed study due to the long time required to finish all simulations and then collect the data that supports this study's results. This tool was developed for the instrumentation and optimization of computational studies conducted by Prof. Tiago A. E. Ferreira's research group.

\section{Results and discussion}
\label{sec:results-and-discussion}

\begin{figure*}[]
\centering
\subfloat[Total Inversion $l = n/N$]{\includegraphics[width=8cm]{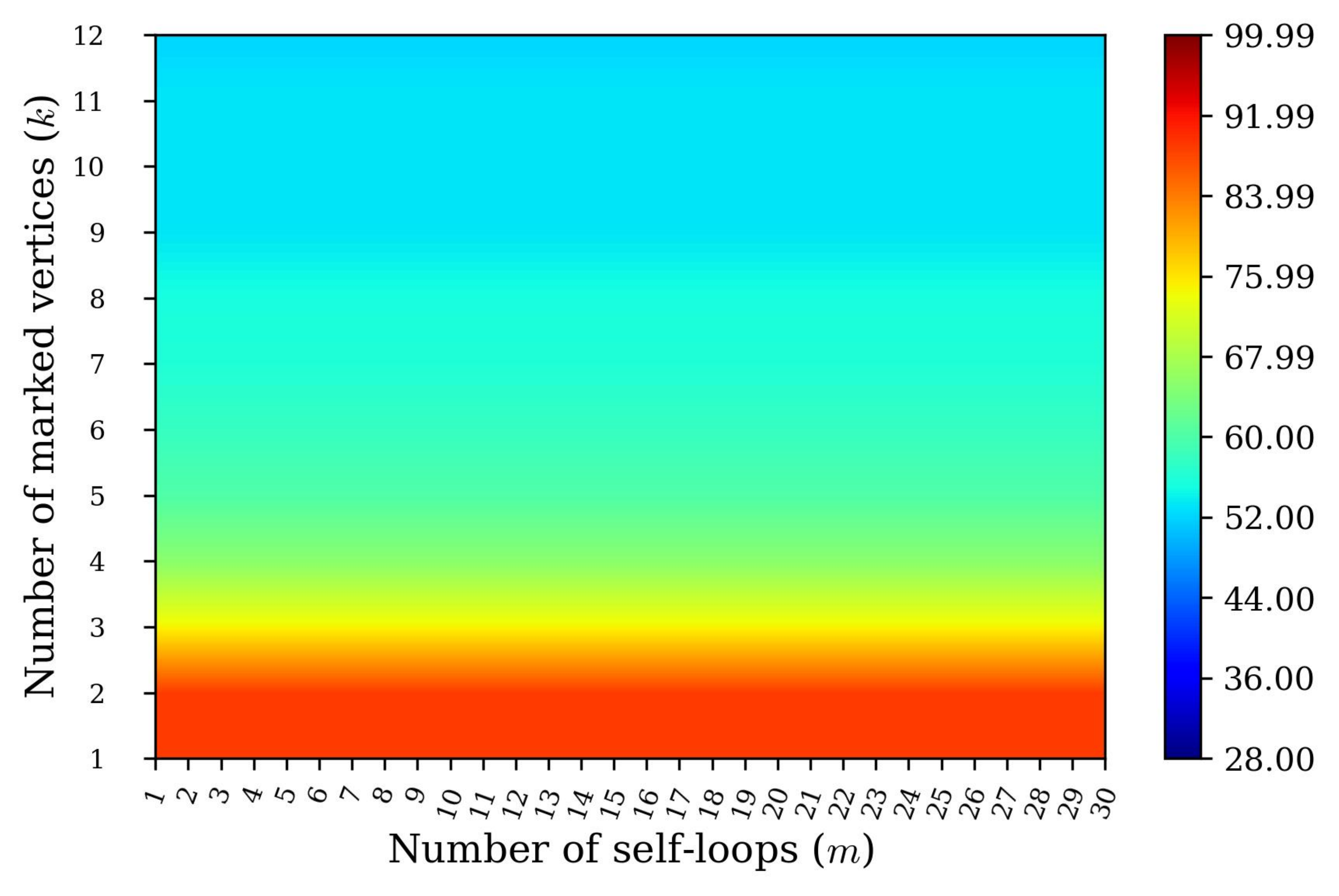}
\label{fig:probability-distribution-non-neighbors-all-signs-n-N}}
\subfloat[Partial Inversion $l = n/N$]{\includegraphics[width=8cm]{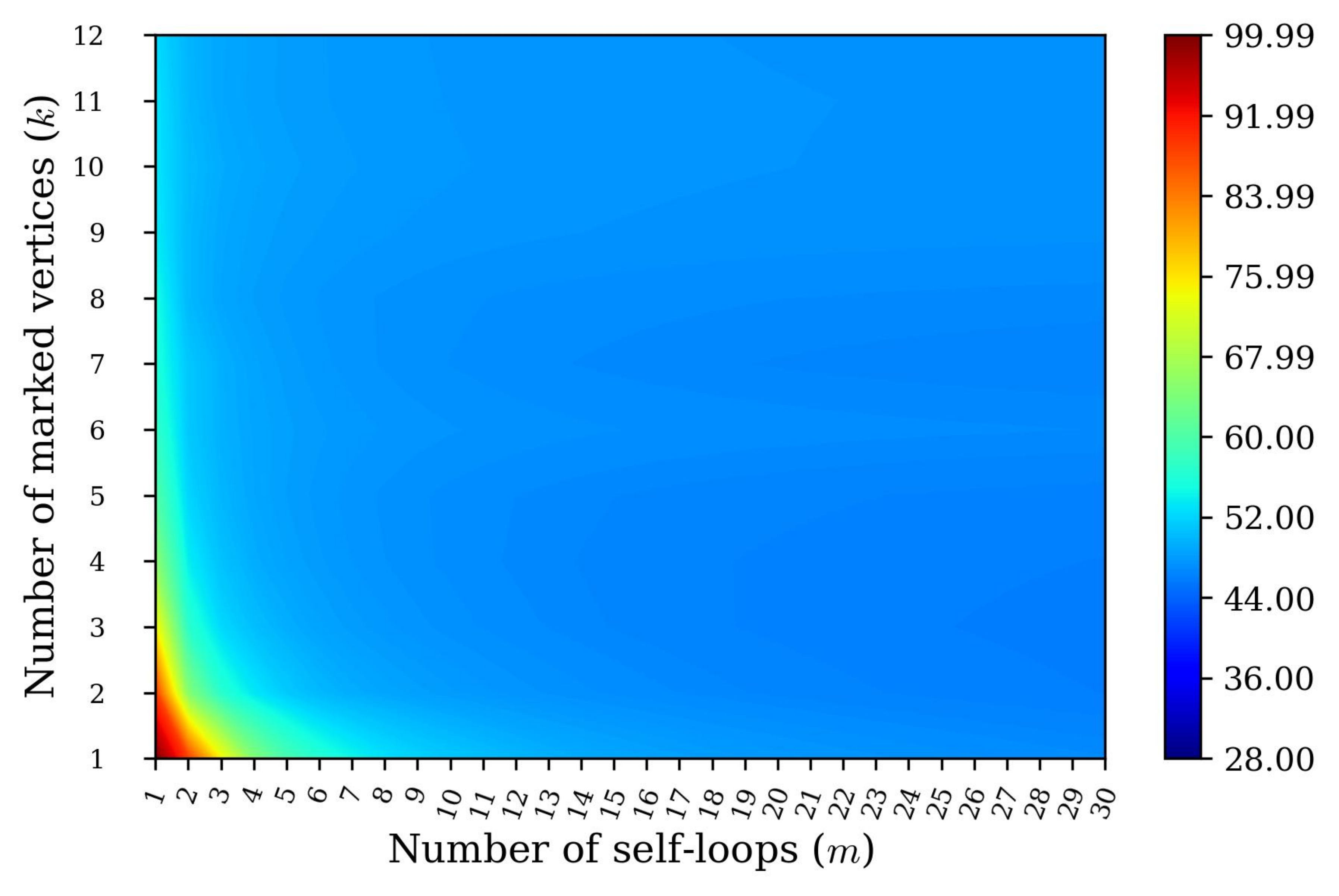}
\label{fig:probability-distribution-non-neighbors-n-N}}\\
\subfloat[Total Inversion $l = (n/N) \cdot k$]{\includegraphics[width=8cm]{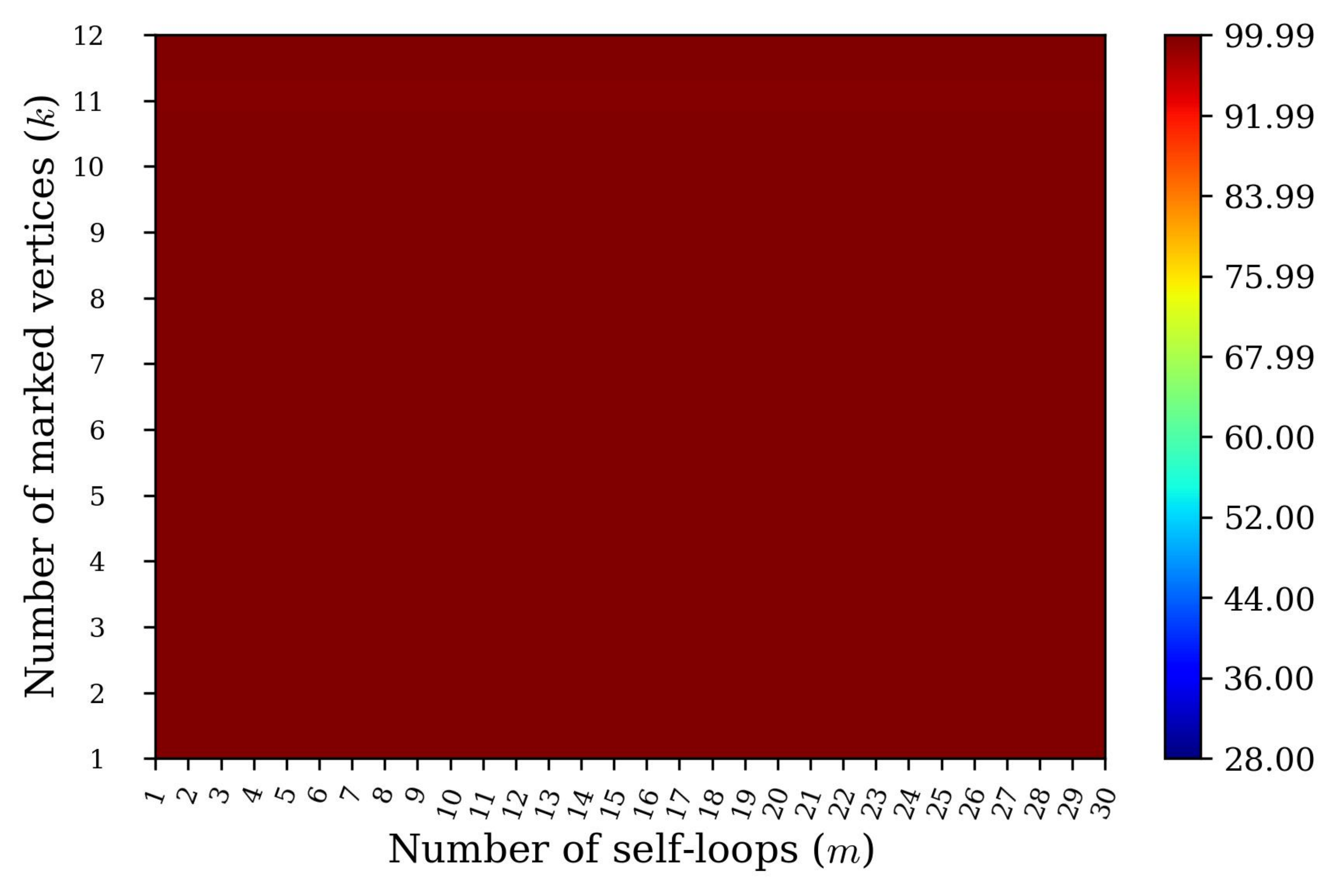}
\label{fig:probability-distribution-non-neighbors-all-signs-n-N-k}}
\subfloat[Partial Inversion $l = (n/N) \cdot k$]{\includegraphics[width=8cm]{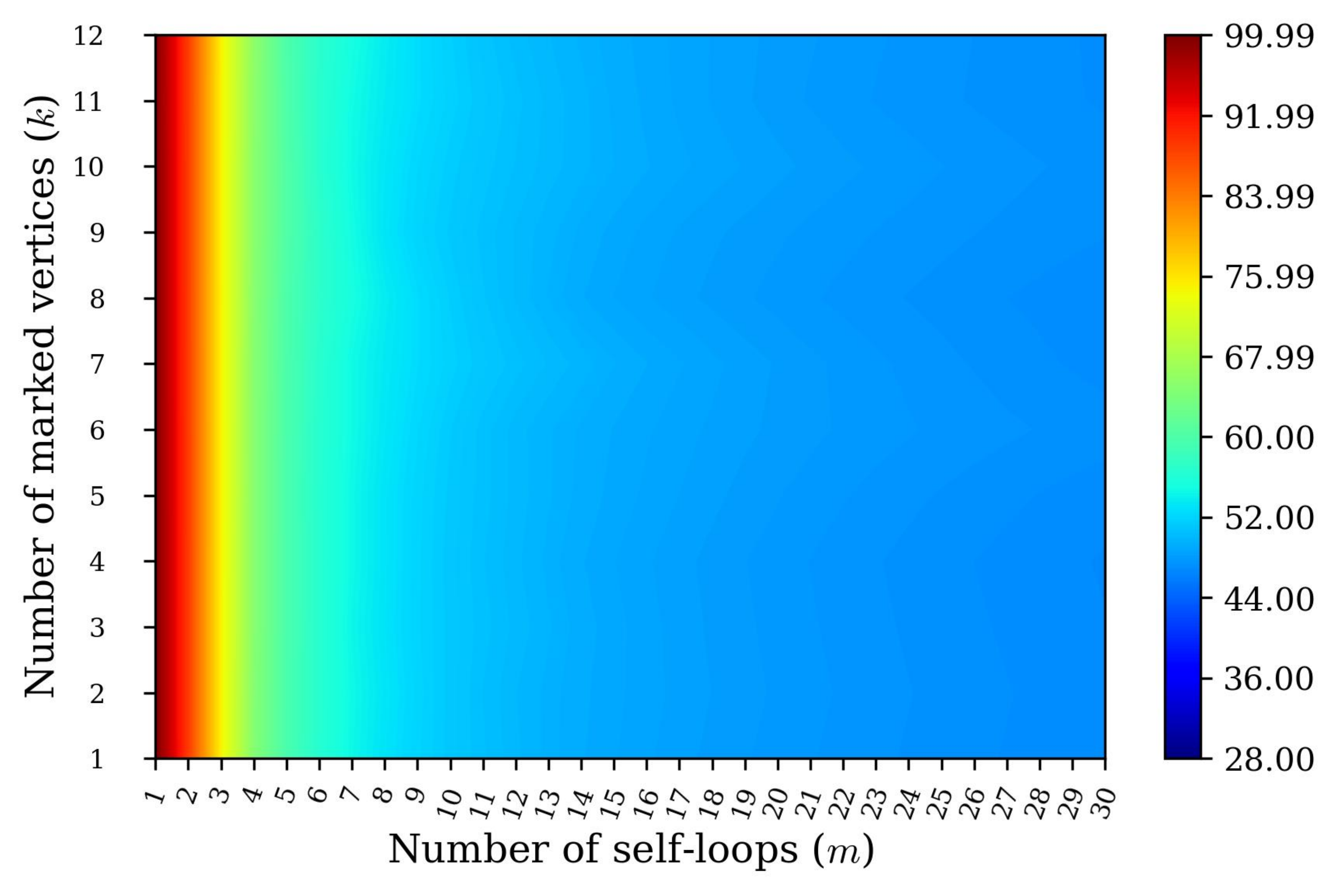}
\label{fig:probability-distribution-non-neighbors-n-N-k}}
\caption{The success probability of the Lackadaisical Quantum Walk with full phase inversion in Figures \protect\subref{fig:probability-distribution-non-neighbors-all-signs-n-N} and \protect\subref{fig:probability-distribution-non-neighbors-all-signs-n-N-k}, and the MSLQW-PPI in Figures \protect\subref{fig:probability-distribution-non-neighbors-n-N} and \protect\subref{fig:probability-distribution-non-neighbors-n-N-k}  in the hypercube to search for non-adjacent marked vertices with $n = 12$ and $N = 4096$ vertices. Figures \protect\subref{fig:probability-distribution-non-neighbors-all-signs-n-N} and \protect\subref{fig:probability-distribution-non-neighbors-n-N} for the weight value $l = n/N$, proposed by \citet{rhodes2020search}. Figures \protect\subref{fig:probability-distribution-non-neighbors-all-signs-n-N-k} and \protect\subref{fig:probability-distribution-non-neighbors-n-N-k} for the weight value $l = (n/N) \cdot k$, proposed by \citet{souza2021lackadaisical}}
\label{fig:probability-distribution-non-neighbors-all-signs-weights-n-N-and-n-N-k}
\end{figure*}

Initially, our discussions focus on cases where only one self-loop is inverted, \textit{i.e.} $s=1$ and $1 \leqslant m \leqslant 30$, and are compared with approaches
where all self-loops are inverted or $s=m$. Fig. \ref{fig:probability-distribution-non-neighbors-all-signs-weights-n-N-and-n-N-k} shows the success probabilities for the lackadaisical quantum walk and the MSLQW-PPI in the hypercube for the weights $l = n/N$ and $l = (n/N) \cdot k$.

Fig.~\ref{fig:probability-distribution-non-neighbors-all-signs-n-N} shows the probability of success for the full phase inversion and Fig.~\ref{fig:probability-distribution-non-neighbors-n-N} shows the probability of success for the partial phase inversion, both for weight values $l = n/N$. When we invert the phase of all self-loops at each marked vertex, it is equivalent to using a single weighted self-loop at each marked vertex on the hypercube \citep{wong2015grover}. In these cases, the results are similar to those obtained in works by \citet{rhodes2020search} with a probability of success of approximately $99\%$ to search for one marked vertex.

\begin{figure*}[]
\centering
\subfloat[Total Inversion $l = n/N$]{\includegraphics[width=8cm]{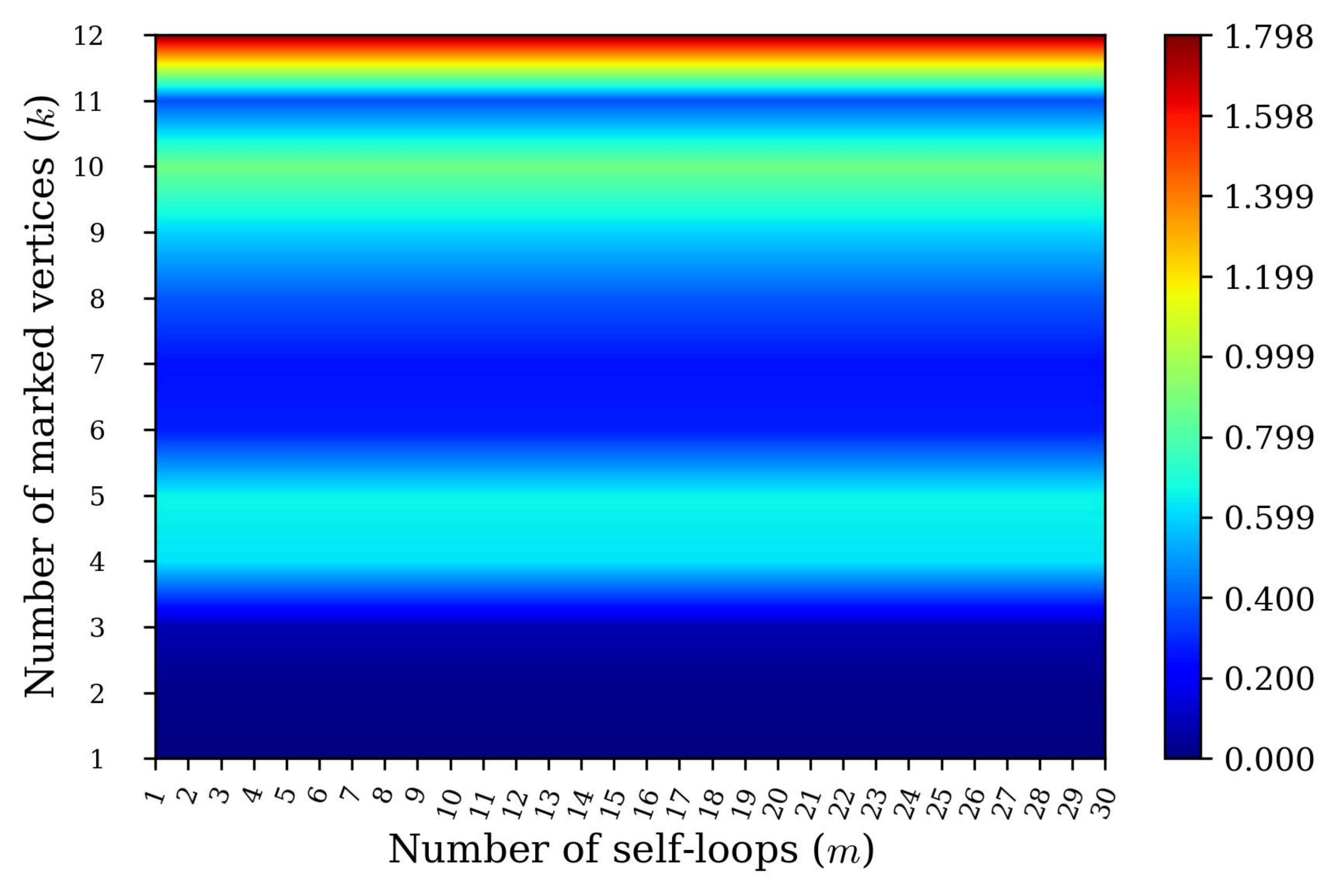}
\label{fig:probability-distribution-non-neighbors-all-signs-n-N-standard-deviation}}
\subfloat[Partial Inversion $l = n/N$]{\includegraphics[width=8cm]{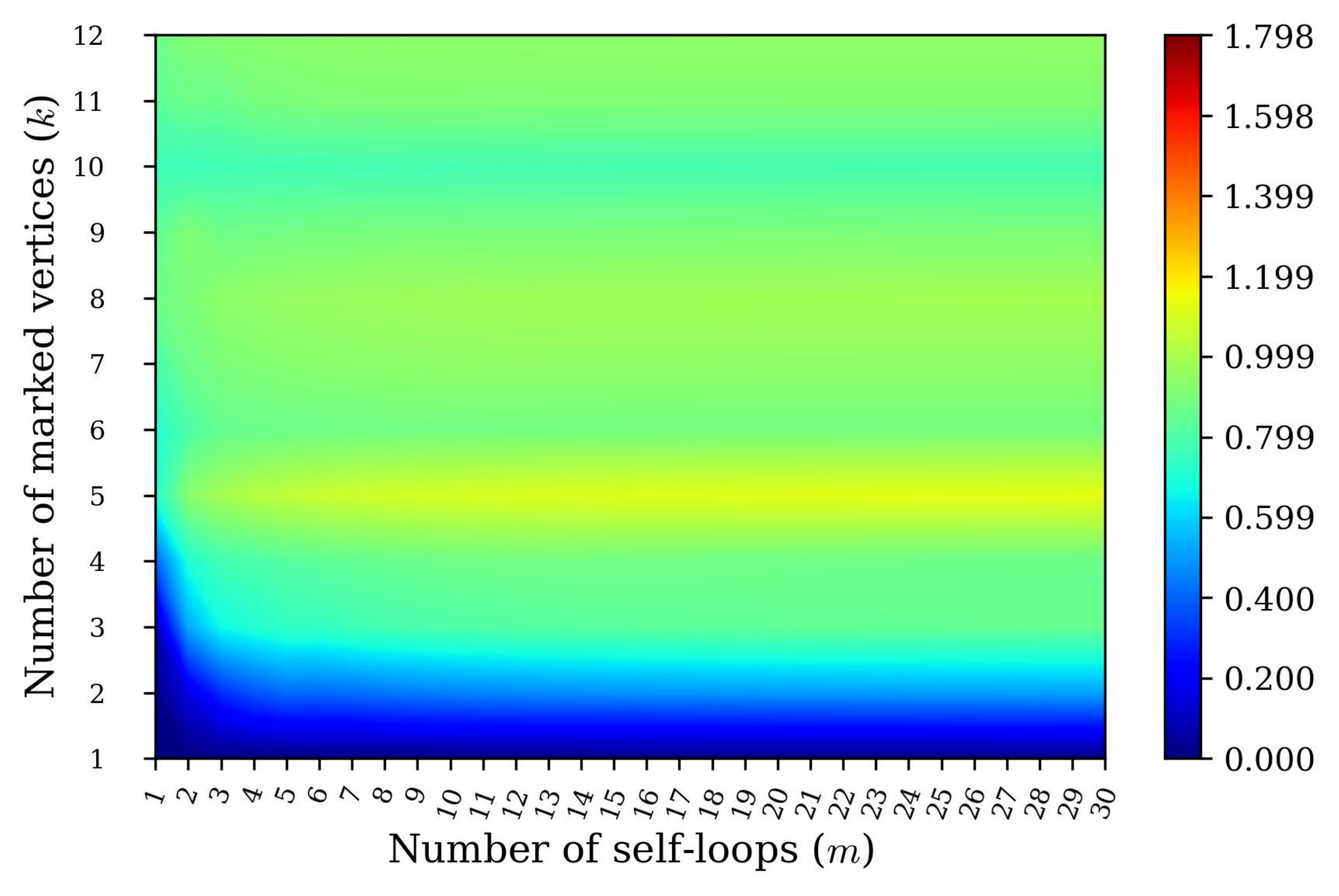}
\label{fig:probability-distribution-non-neighbors-n-N-standard-deviation}}\\
\subfloat[Total Inversion $l = (n/N) \cdot k$]{\includegraphics[width=8cm]{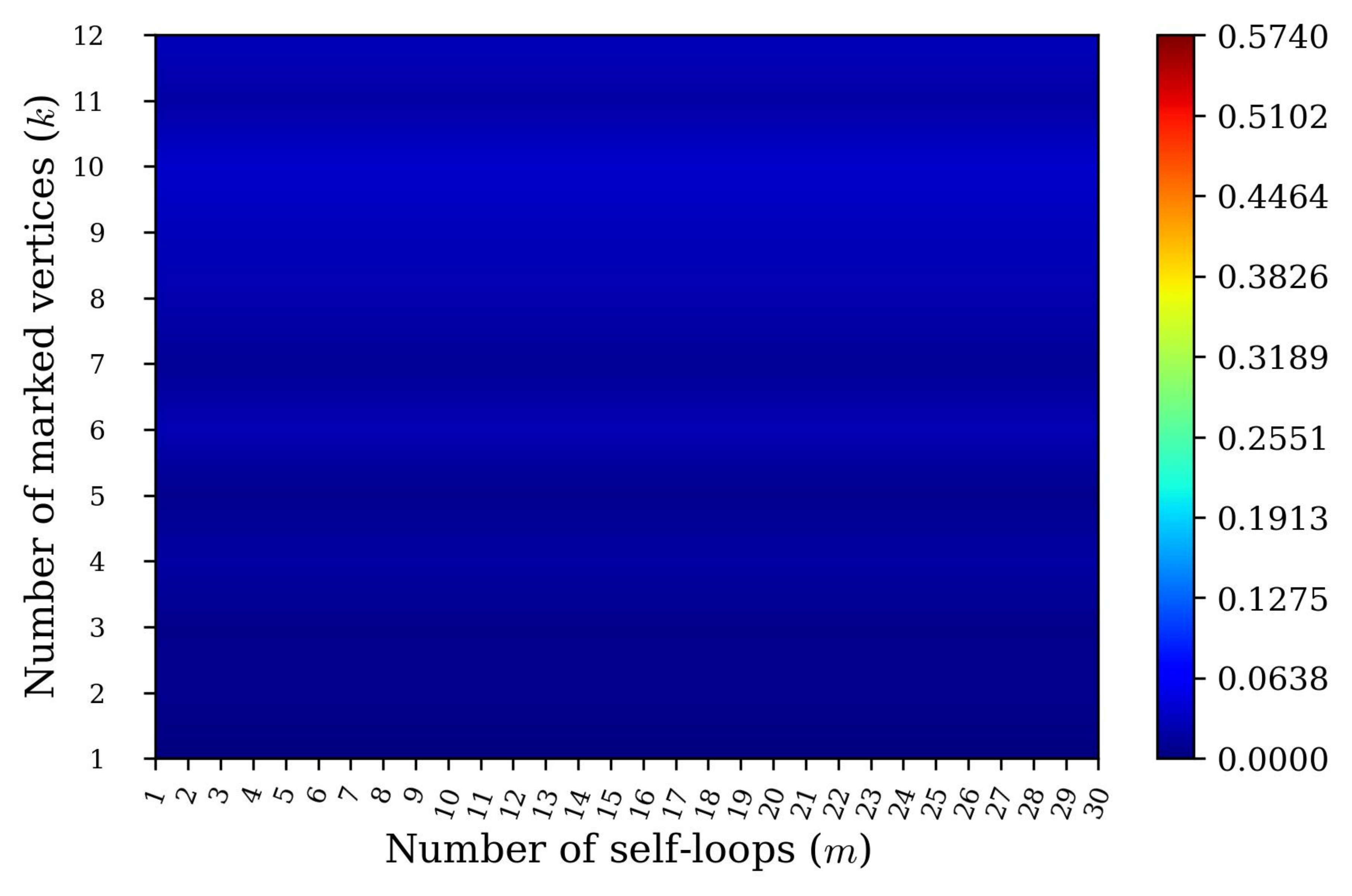}
\label{fig:probability-distribution-non-neighbors-all-signs-n-N-k-standard-deviation}}
\subfloat[Partial Inversion $l = (n/N) \cdot k$]{\includegraphics[width=8cm]{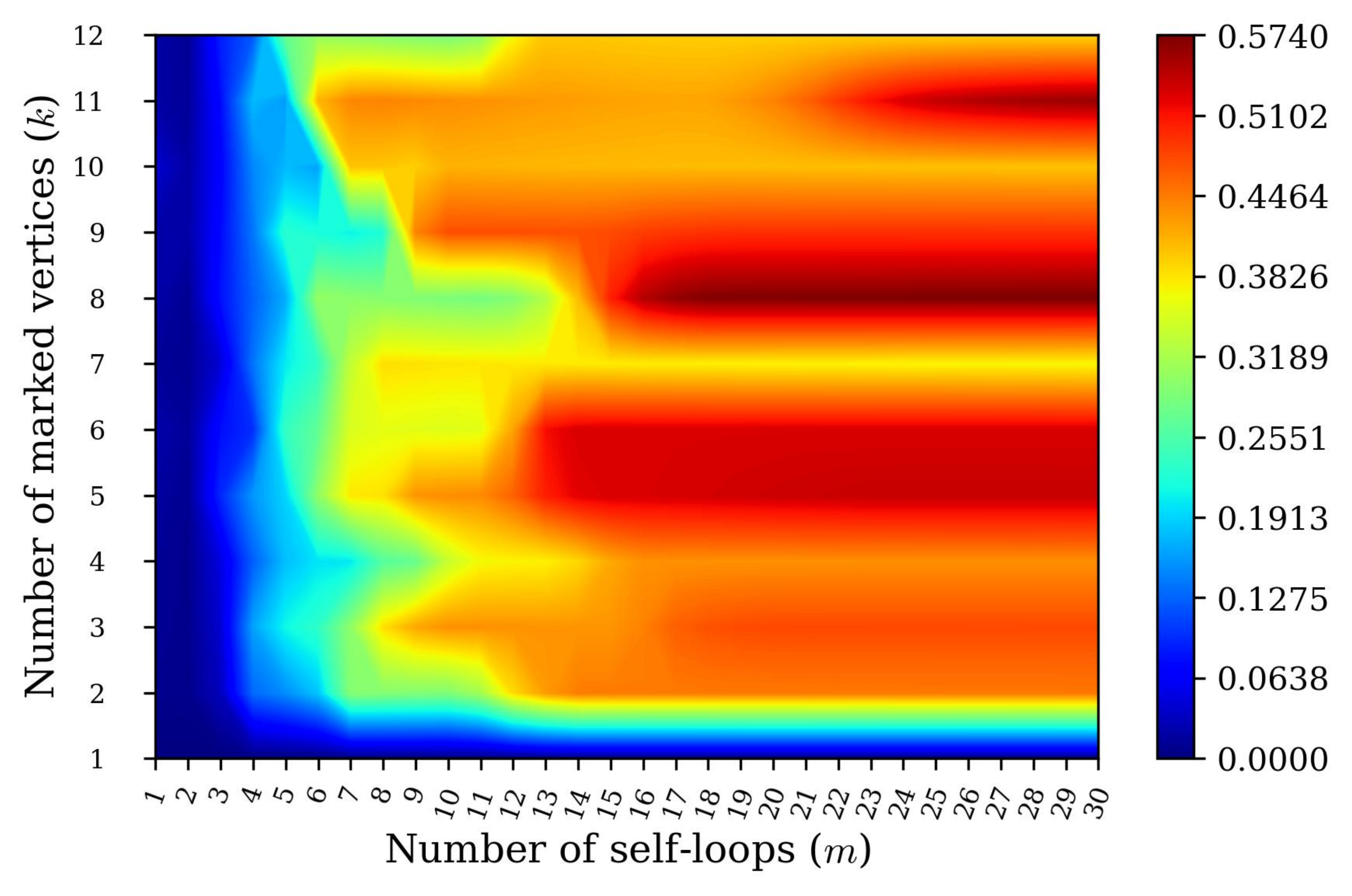}
\label{fig:probability-distribution-non-neighbors-n-N-k-standard-deviation}}
\caption{The results for coefficients of variation. Their values are represented in percentage terms. \protect\subref{fig:probability-distribution-non-neighbors-all-signs-n-N-standard-deviation} and \protect\subref{fig:probability-distribution-non-neighbors-n-N-standard-deviation} weight value $l = n/N$. \protect\subref{fig:probability-distribution-non-neighbors-all-signs-n-N-k-standard-deviation} and \protect\subref{fig:probability-distribution-non-neighbors-n-N-k-standard-deviation} weight value $l = (n/N) \cdot k$.}
\label{fig:cv-non-neighbors-n-N-and-n-N-k}
\end{figure*}

Fig.~\ref{fig:probability-distribution-non-neighbors-all-signs-n-N-k} shows the probability of success for the full phase inversion and Fig.~\ref{fig:probability-distribution-non-neighbors-n-N-k} shows the probability of success for the partial phase inversion, both for weight values $l = (n/N) \cdot k$. In Fig. \ref{fig:probability-distribution-non-neighbors-all-signs-n-N-k}, where the full phase inversion occurs, the probability of success stays at values close to $1$, even using multiple self-loops, with a small variation if we observe its coefficient of variation in Fig. \ref{fig:probability-distribution-non-neighbors-all-signs-n-N-k-standard-deviation}. As we can see in Fig. \ref{fig:probability-distribution-non-neighbors-n-N-k}, the probabilities of success present results similar to those obtained by \citet{souza2021lackadaisical}.

Fig. \ref{fig:cv-non-neighbors-n-N-and-n-N-k} shows the coefficient of variation ($\sigma_w / \bar{w}$ -- standard deviation normalized by the mean value)  to measure the level of dispersion presented in the results obtained with the total and partial phase inversion of the results presented in Fig \ref{fig:probability-distribution-non-neighbors-all-signs-weights-n-N-and-n-N-k}. The level of dispersion presented in the results obtained with the partial phase inversion in Figures \ref{fig:probability-distribution-non-neighbors-n-N-standard-deviation} and \ref{fig:probability-distribution-non-neighbors-n-N-k-standard-deviation} describes a behavior where the maximum probabilities of success present a smaller coefficient of variation. Unlike the results presented in Fig.~\ref{fig:probability-distribution-non-neighbors-all-signs-n-N-standard-deviation} obtained with full inversion where the level of dispersion presented is greater for lower probabilities. Analyzing Fig. \ref{fig:probability-distribution-non-neighbors-all-signs-n-N-k-standard-deviation}, although the result for the coefficient of variation appears stable, there are subtle variations concerning the maximum probabilities of success presented in Fig.~\ref{fig:probability-distribution-non-neighbors-all-signs-n-N-k}.

As we see in Fig.~\ref{fig:probability-distribution-non-neighbors-all-signs-weights-n-N-and-n-N-k}, the weight values proposed by \citet{rhodes2020search} and \citet{souza2021lackadaisical} are ideal to use in approaches with a single self-loop at each vertex. Therefore, for our approach, it is necessary to define the best weight value for using multiple self-loops. Most of the weights defined in previous works for structures such as a complete graph, hypercube, one-dimensional grid, and two-dimensional grid consider three main parameters: the vertex degree, the total number of vertices, and the number of marked vertices. Two weight values were already defined by \citet{rhodes2020search} and \citet{souza2021lackadaisical} as ideal for searching for one and multiple marked vertices in the hypercube, respectively: $l = n/N$ and $l = (n/N)\cdot k$, where $n$ is the degree of the vertices, $N$ the number of vertices, and $k$ the number of marked vertices.

\begin{figure*}[]
\centering
\subfloat[Total Inversion $l = n^{2}/N$]{\includegraphics[width=8cm]{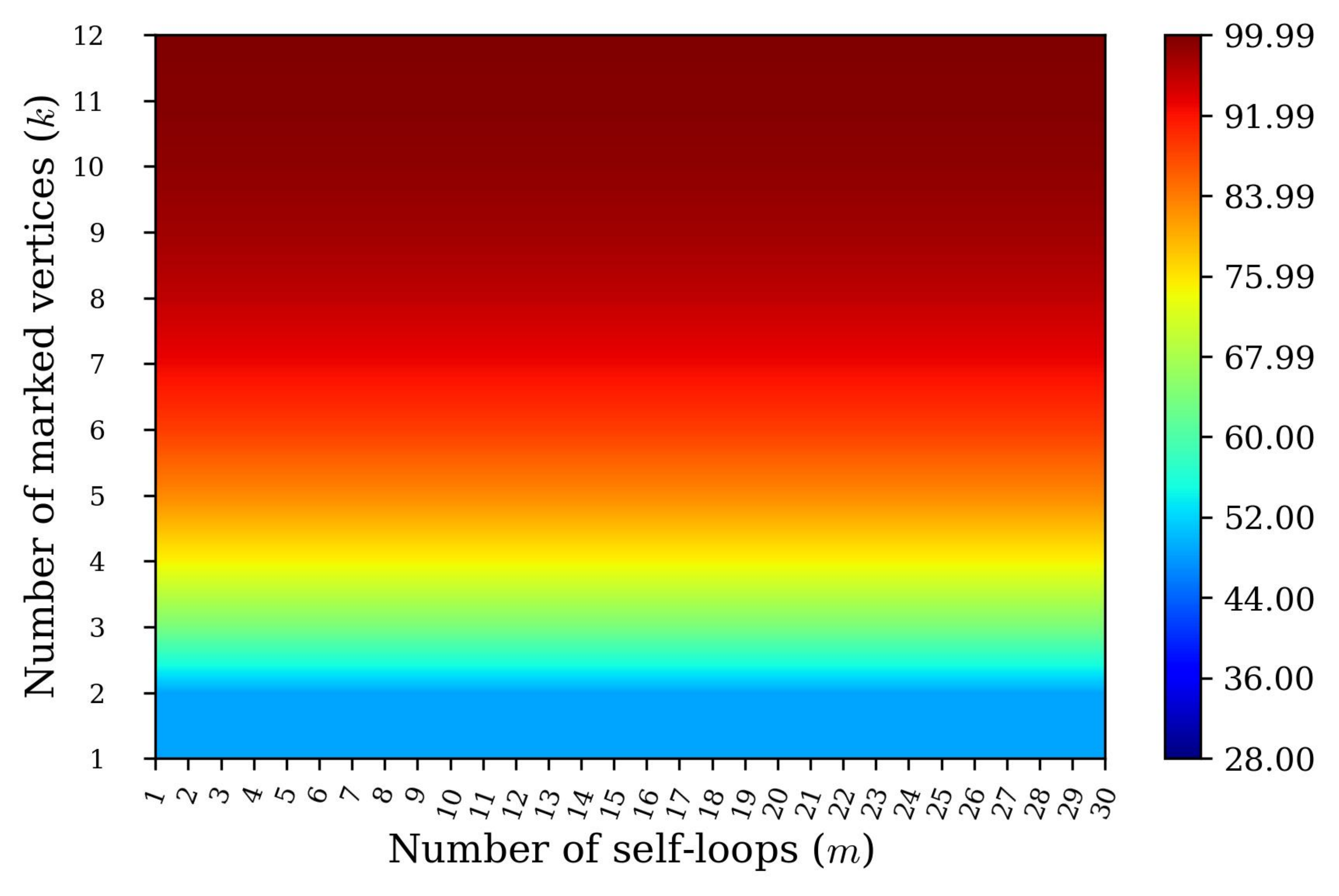}
\label{fig:probability-distribution-non-neighbors-all-signs-n-2-N}}
\subfloat[Partial Inversion $l = n^{2}/N$]{\includegraphics[width=8cm]{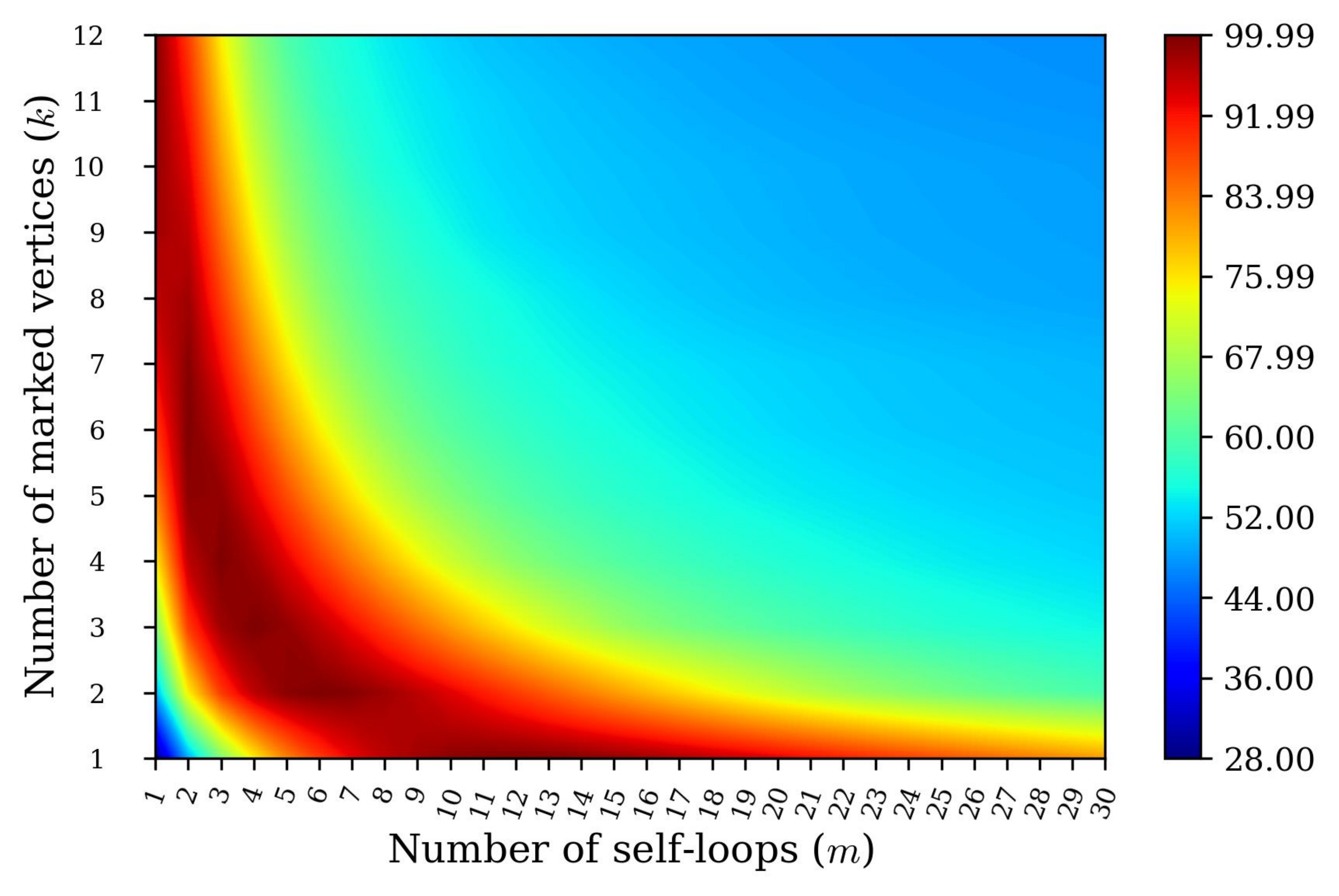}
\label{fig:probability-distribution-non-neighbors-n-2-N}}\\
\subfloat[Total Inversion $l = (n^{2}/N) \cdot k$]{\includegraphics[width=8cm]{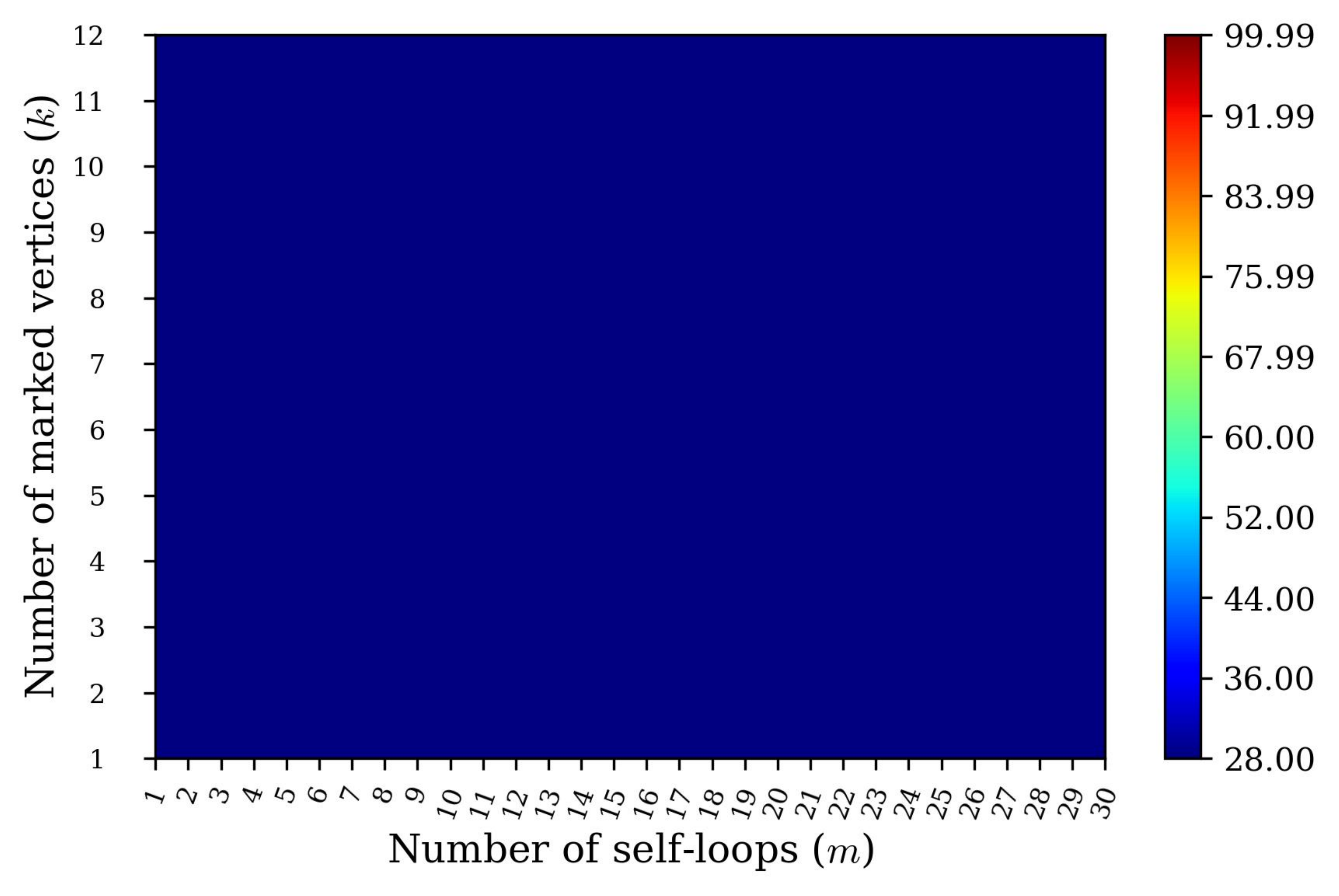}
\label{fig:probability-distribution-non-neighbors-all-signs-n-2-N-k}}
\subfloat[Partial Inversion $l = (n^{2}/N) \cdot k$]{\includegraphics[width=8cm]{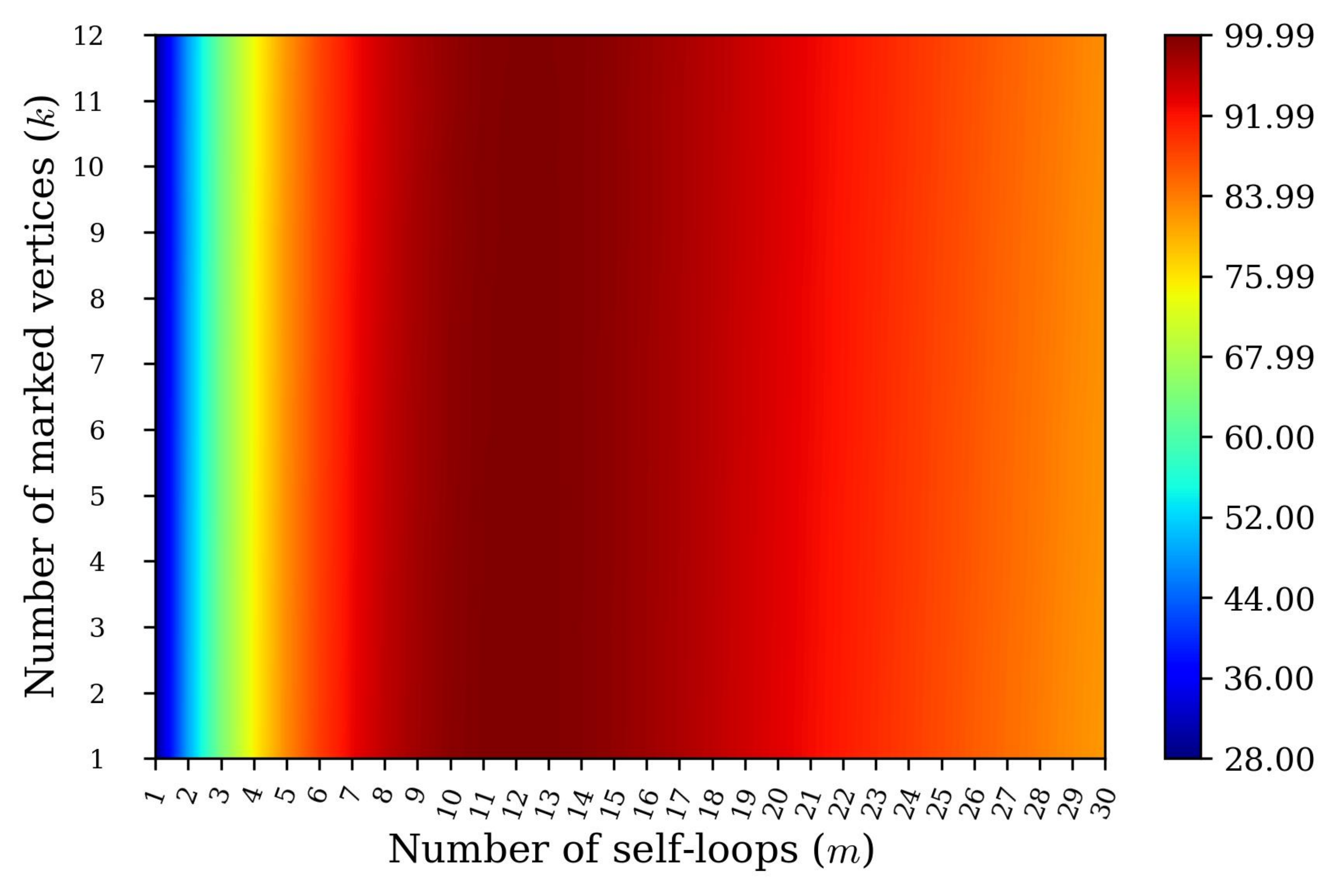}
\label{fig:probability-distribution-non-neighbors-n-2-N-k}}
\caption{The success probability of the Lackadaisical Quantum Walk with full phase inversion in Figures \protect\subref{fig:probability-distribution-non-neighbors-all-signs-n-2-N} and \protect\subref{fig:probability-distribution-non-neighbors-all-signs-n-2-N-k}, and the MSLQW-PPI in Figures \protect\subref{fig:probability-distribution-non-neighbors-n-2-N} and \protect\subref{fig:probability-distribution-non-neighbors-n-2-N-k} in the hypercube to search for non-adjacent marked vertices with $n = 12$ and $N = 4096$ vertices. Figures \protect\subref{fig:probability-distribution-non-neighbors-all-signs-n-2-N} and \protect\subref{fig:probability-distribution-non-neighbors-n-2-N} for the weight value $l = n^{2}/N$. Figures \protect\subref{fig:probability-distribution-non-neighbors-all-signs-n-2-N-k} and \protect\subref{fig:probability-distribution-non-neighbors-n-2-N-k} for the weight value $l = (n^{2}/N) \cdot k$.}
\label{fig:probability-distribution-non-neighbors-all-signs-weights-n-2-N-and-n-2-N-k}
\end{figure*}

The idea here is to propose two new weight values based on those previous ideal weight values used to search for one and multiple vertices in the hypercube. We introduce a new parameter in the weight composition: the exponent in the numerator. Considering that the weights $l = n/N$ and $l = (n/N)\cdot k$ have an exponent equal to $1$ in the parameter $n^{1}$, we can consider other values for this exponent. We can modify the scale of weight values through this new exponent. So we have a new proposal for the two weights

\begin{align*}
    l = \left ( \frac{n^{2}}{N}\right ), \text{ and } l = \left ( \frac{n^{2}}{N}\right )\cdot k.
\end{align*}
Fig.~\ref{fig:probability-distribution-non-neighbors-all-signs-weights-n-2-N-and-n-2-N-k} shows the probability of success of the lackadaisical quantum walk with full phase inversion and MSLQW-PPI using the two new weight values to search multiple marked vertices. 
More one time, the first column (Figures~\ref{fig:probability-distribution-non-neighbors-all-signs-n-2-N} and~\ref{fig:probability-distribution-non-neighbors-all-signs-n-2-N-k}) presents the results for the full phase inversion and the second column (\ref{fig:probability-distribution-non-neighbors-n-2-N} and \ref{fig:probability-distribution-non-neighbors-n-2-N-k}) the results for the partial phase inversion to the simplest situation of $s=1$ and $1 \leqslant m \leqslant 30$.

Figures~\ref{fig:probability-distribution-non-neighbors-all-signs-n-2-N} and~\ref{fig:probability-distribution-non-neighbors-n-2-N} show the maximum probability of success for the weight value $l = n^{2}/N$. In the Fig.~\ref{fig:probability-distribution-non-neighbors-all-signs-n-2-N}, the phase of all components of the target state representing the $m$ self-loops is inverted. In this case, the probability of success is affected reaching a maximum probability of approximately $p \approx 1$ only when $k = 11, 12$. The Fig.~\ref{fig:probability-distribution-non-neighbors-n-2-N} shows the maximum probability of success for the partial inversion. In this case, we obtain success probabilities close to $1$. For $k=2,3,4$ the maximum success probability is reached (with $m=6,4,3$) self-loops, respectively, and for $k=5,6,7,8$ (with $m=2$). Fig.~\ref{fig:probability-distribution-non-neighbors-all-signs-n-2-N-k} shows the maximum probability of success where the phase of all components of the target state is inverted and Fig.~\ref{fig:probability-distribution-non-neighbors-n-2-N-k} shows the maximum probability of success for the partial inversion, both for the weight value $l = (n^{2}/N)\cdot k$. This scenario showed a considerable gain in the maximum probability of success from $p = 0.28$ (with $\text{m}=1$) to $p \approx 1$ (with $\text{m} = 12$) self-loops to any number $k$ of the marked vertices.

\begin{figure*}[]
\centering
\subfloat[Total Inversion $l = n^{2}/N$]{\includegraphics[width=8cm]{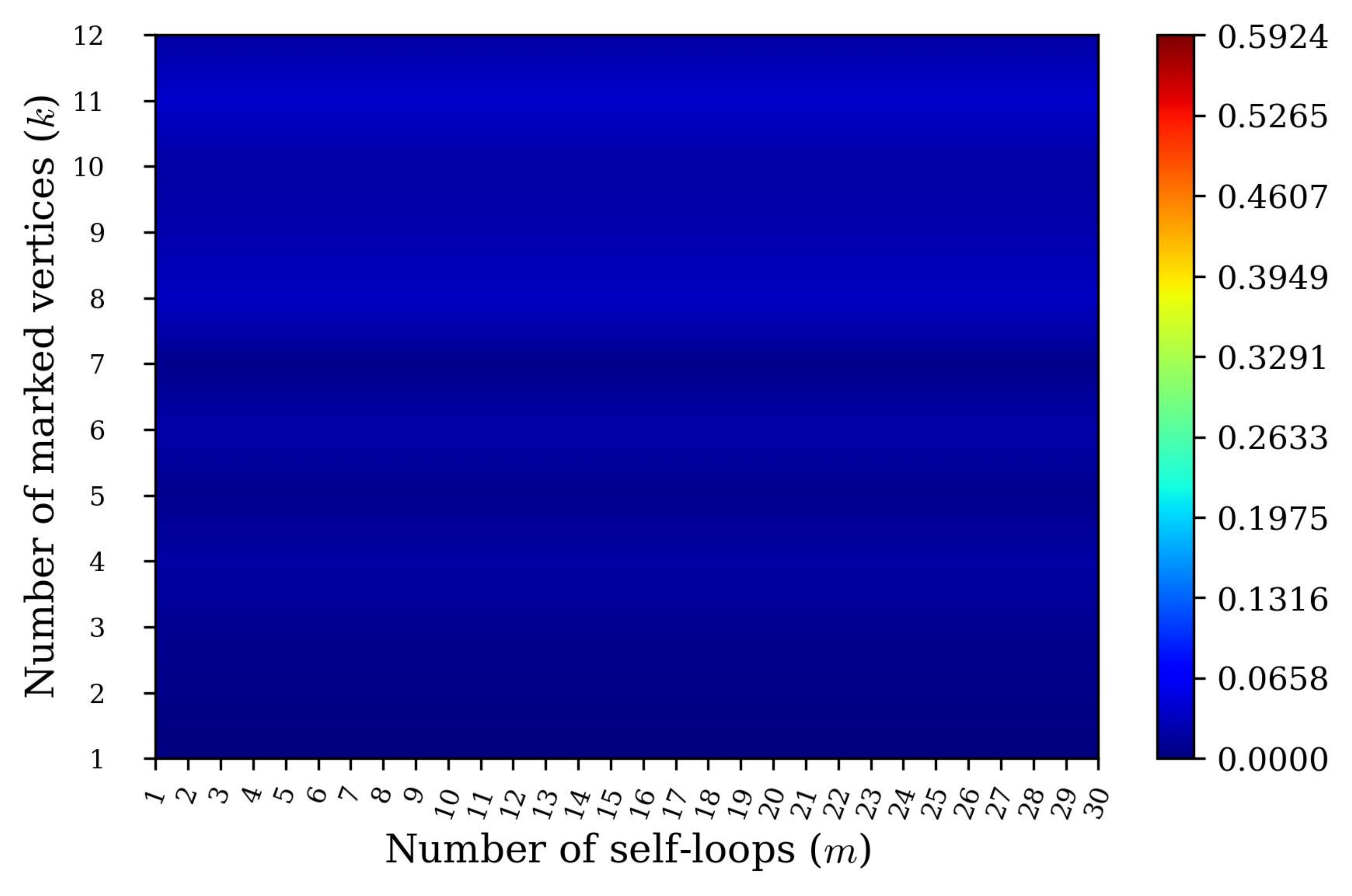}
\label{fig:n-2-N-non-neighbors-standard-deviation-all-phase-invertion}}
\subfloat[Partial Inversion $l = n^{2}/N$]{\includegraphics[width=8cm]{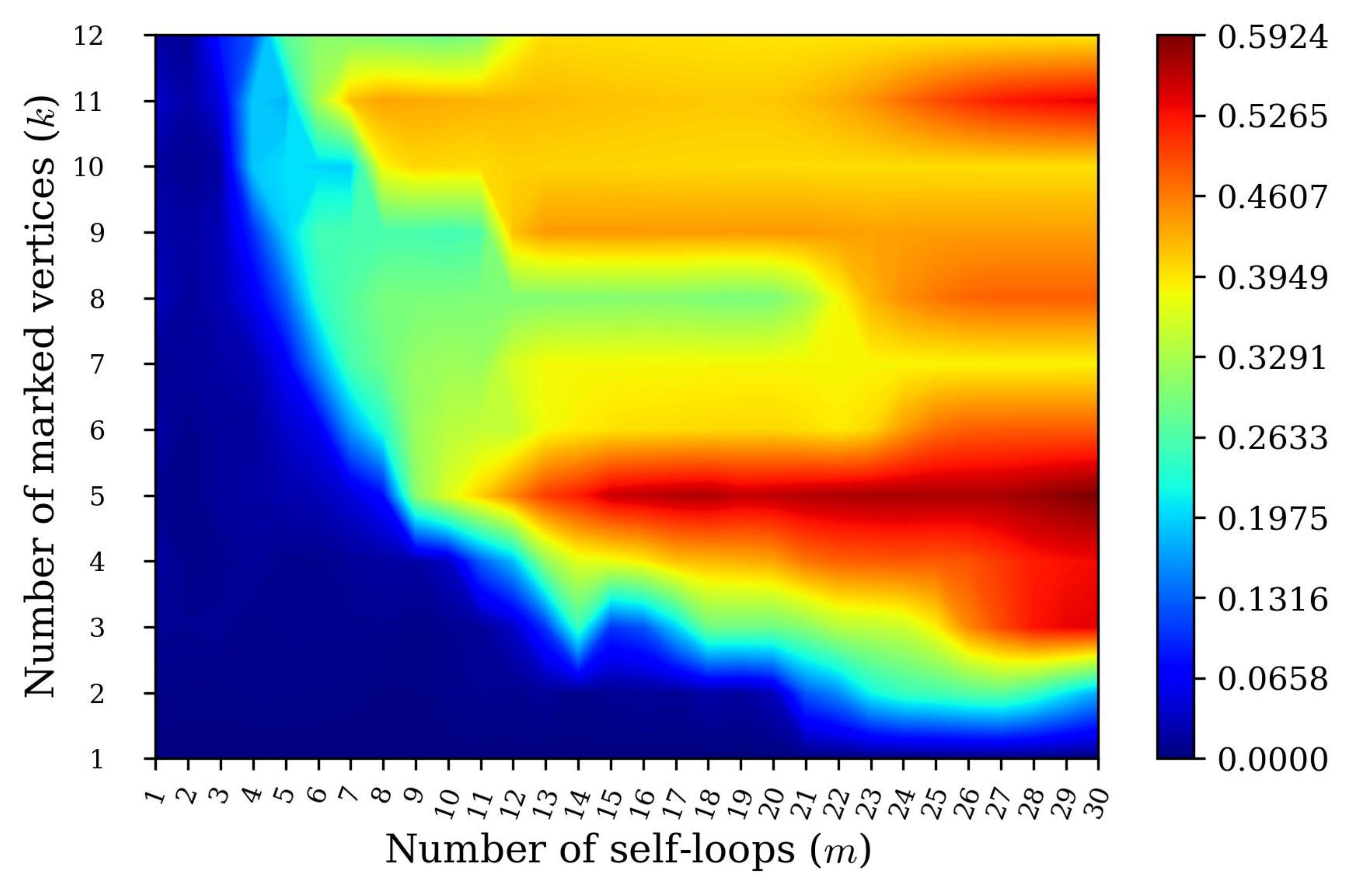}
\label{fig:probability-distribution-non-neighbors-n-2-N-standard-deviation}}
\\
\subfloat[Total Inversion $l = (n^{2}/N) \cdot k$]{\includegraphics[width=8cm]{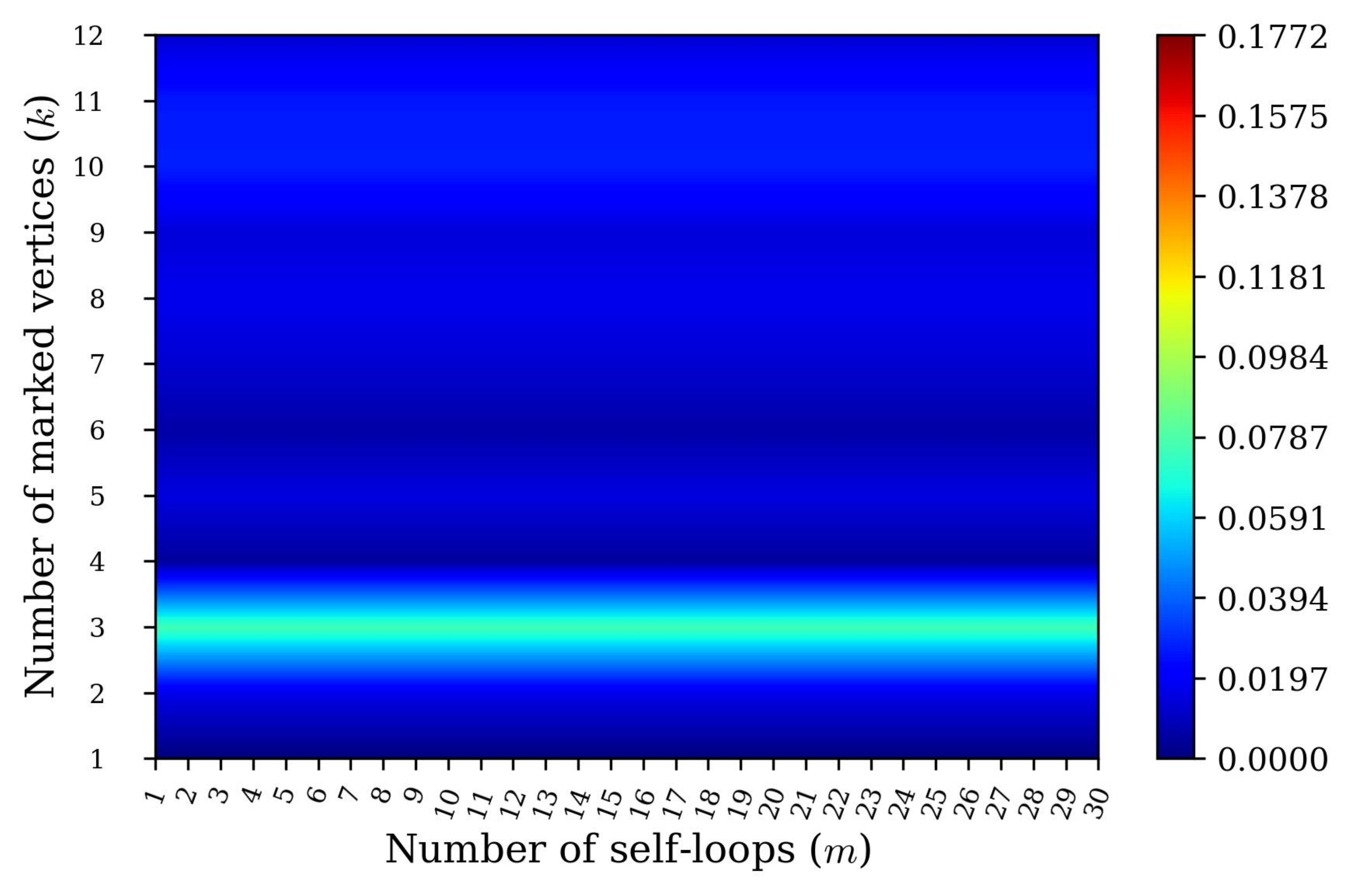}
\label{fig:n-2-N-k-non-neighbors-standard-deviation-all-phase-invertion}}
\subfloat[Partial Inversion $l = (n^{2}/N) \cdot k$]{\includegraphics[width=8cm]{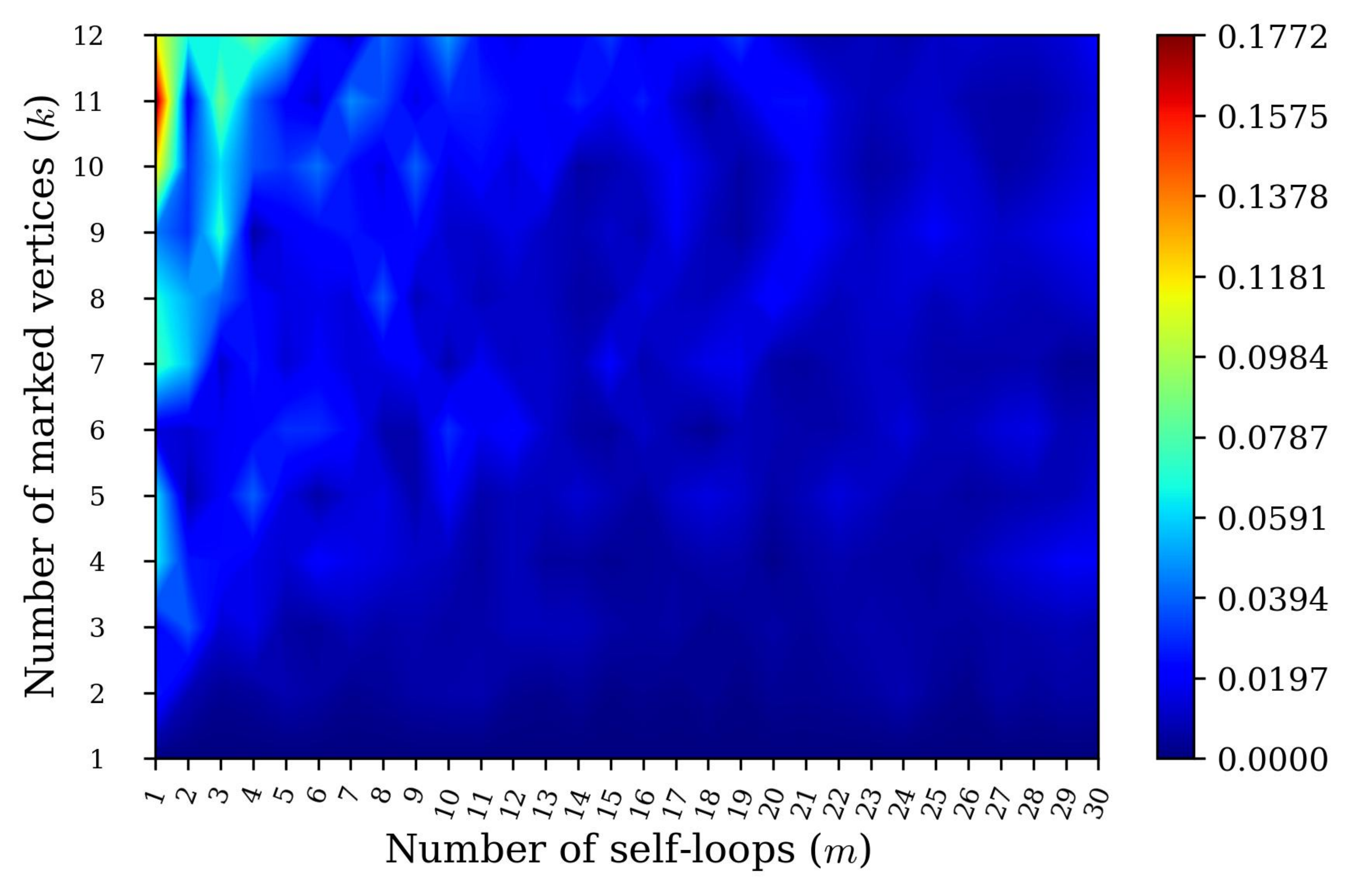}
\label{fig:probability-distribution-non-neighbors-n-2-N-k-standard-deviation}}
\caption{The results for coefficients of variation. Their values are represented in percentage terms. \protect\subref{fig:probability-distribution-non-neighbors-all-signs-n-N-standard-deviation} and \protect\subref{fig:probability-distribution-non-neighbors-n-N-standard-deviation} weight value $l = n^{2}/N$. \protect\subref{fig:n-2-N-k-non-neighbors-standard-deviation-all-phase-invertion} and \protect\subref{fig:probability-distribution-non-neighbors-n-N-k-standard-deviation} weight value $l = (n^{2}/N) \cdot k$.}
\label{fig:probability-distribution-non-neighbors-n-2-N-and-n-2-N-k}
\end{figure*}

Figure \ref{fig:probability-distribution-non-neighbors-n-2-N-and-n-2-N-k} shows the coefficient of variation for the results described in Fig.~\ref{fig:probability-distribution-non-neighbors-all-signs-weights-n-2-N-and-n-2-N-k}. Analyzing the results, we observe a different behavior between the complete and partial phase inversions. With the partial phase inversion, the small coefficient of variation coincides with the maximum probabilities of success, however, with the complete phase inversion the variation is also smaller but does not coincide with the maximum probabilities of success. Table \ref{tab:comp-weight-and-self-loops-all-signs-n-2-N} shows that in most of the results applying partial phase inversion, there was an improvement in the maximum probability of success and a smaller coefficient of variation.

\begin{table*}[]
\centering
\caption{Comparison between the probability of success and number of self-loops for two scenarios. The column for Fig.~\ref{fig:probability-distribution-non-neighbors-all-signs-n-2-N} represents the results where the phase of all self-loops is reversed. The column for Fig.~\ref{fig:probability-distribution-non-neighbors-n-2-N} is for the case where the phase of only one self-loop is inverted. Both using the weight $l = n^{2}/N$. The acronym cv means the coefficient of variation.}
\label{tab:comp-weight-and-self-loops-all-signs-n-2-N}
\begin{tabular}{lcccccc}
\toprule
 & \multicolumn{6}{c}{Figures} \\ \cmidrule{2-7} 
 & \multicolumn{3}{c}{\ref{fig:probability-distribution-non-neighbors-all-signs-n-2-N}} & \multicolumn{3}{c}{\ref{fig:probability-distribution-non-neighbors-n-2-N}} \\ \cmidrule(lr){2-4}\cmidrule(lr){5-7}
k & p    & $m$ & cv           & p    & $m$ & cv \\ \midrule
2 & 0.48 & 1 & 5.234e-05 & 0.99 & 6 & 3.536e-04 \\
3 & 0.64 & 1 & 1.050e-04 & 0.99 & 4 & 2.177e-04 \\
4 & 0.75 & 1 & 1.922e-04 & 0.99 & 3 & 9.214e-05 \\
5 & 0.83 & 1 & 9.214e-05 & 0.99 & 2 & 1.922e-04 \\
6 & 0.88 & 1 & 2.177e-04 & 0.99 & 2 & 1.922e-04 \\
7 & 0.92 & 1 & 8.130e-05 & 0.99 & 2 & 1.922e-04 \\
8 & 0.95 & 1 & 3.536e-04 & 0.97 & 2 & 1.922e-04 \\
9 & 0.97 & 1 & 2.477e-04 & 0.97 & 1 & 1.050e-04 \\ \bottomrule
\end{tabular}
\end{table*}
Continuing the comparisons between the results obtained with the different weight values, again, we observed that it was possible to increase the probability of success with more than one self-loop and partial phase inversion. Fig.~\ref{fig:probability-distribution-non-neighbors-n-N-k} shows the probability of success for the weight $l = (n/N)\cdot k$ with total phase inversion. Compared to the results presented in Fig.~\ref{fig:probability-distribution-non-neighbors-n-2-N-k} that shows the probability of success for the weight $l = (n^{2}/N)\cdot k$ with partial phase inversion, the probabilities of success are similar and close to $1$, and what differs is the number of self-loops per vertex. In practical terms, using a single self-loop in this scenario is better. However, it is important to note that there exists a relationship between the weight and the number of self-loops per vertex. Note that the number of self-loops may change depending on the weight value. In the case of Fig.~\ref{fig:probability-distribution-non-neighbors-n-2-N-k}, the probability of success is maximized when the number of self-loops increases to 12.

Finally, let us compare the results obtained from the partial inversion using all four weights. The results presented in Figures~\ref{fig:probability-distribution-non-neighbors-n-N} and~\ref{fig:probability-distribution-non-neighbors-n-N-k} show that the weights proposed by \citet{rhodes2020search} and \citet{souza2021lackadaisical}, $l = n/N$ and $l = (n/N)\cdot k$ are not the ideal weights for MSLQW - PPI. However, with the use of the weights proposed in this work, $l = n^{2}/N$ and $l = (n^{2}/N)\cdot k$, we achieved the best results. We highlight the weight value $l = (n^{2}/N)\cdot k$. With this weight value, we obtained a stable behavior in most of the results. Table \ref{tab:comparison-weights} shows the results of this comparison for the probability of success, number of self-loops, and coefficient of variation for all weights used in the MSLQW - PPI. Note that by modifying the weight scale and using various self-loops and partial phase inversion, it is possible to increase the probability of success.

Now, our discussions focus on cases where more than one self-loop is inverted. Preliminary results indicate that the phase inversion of a single self-loop $\circlearrowleft_{\tau=j}$ is sufficient to obtain the results presented in this work. However, the results also showed that it is possible to achieve maximum success probabilities close to $1$ by inverting $1 < s < m$ self-loops.

The results of the experiments show that as the number of inverted self-loops $s$ increases, the number of self-loops $m$ needed to achieve maximum success probabilities $p \approx 1$ also increases. However, it is also possible to find the required number of self-loops $m$ to achieve maximum success probabilities close to $1$. 

\begin{figure*}[]
    \centering
    \includegraphics[width=8cm]{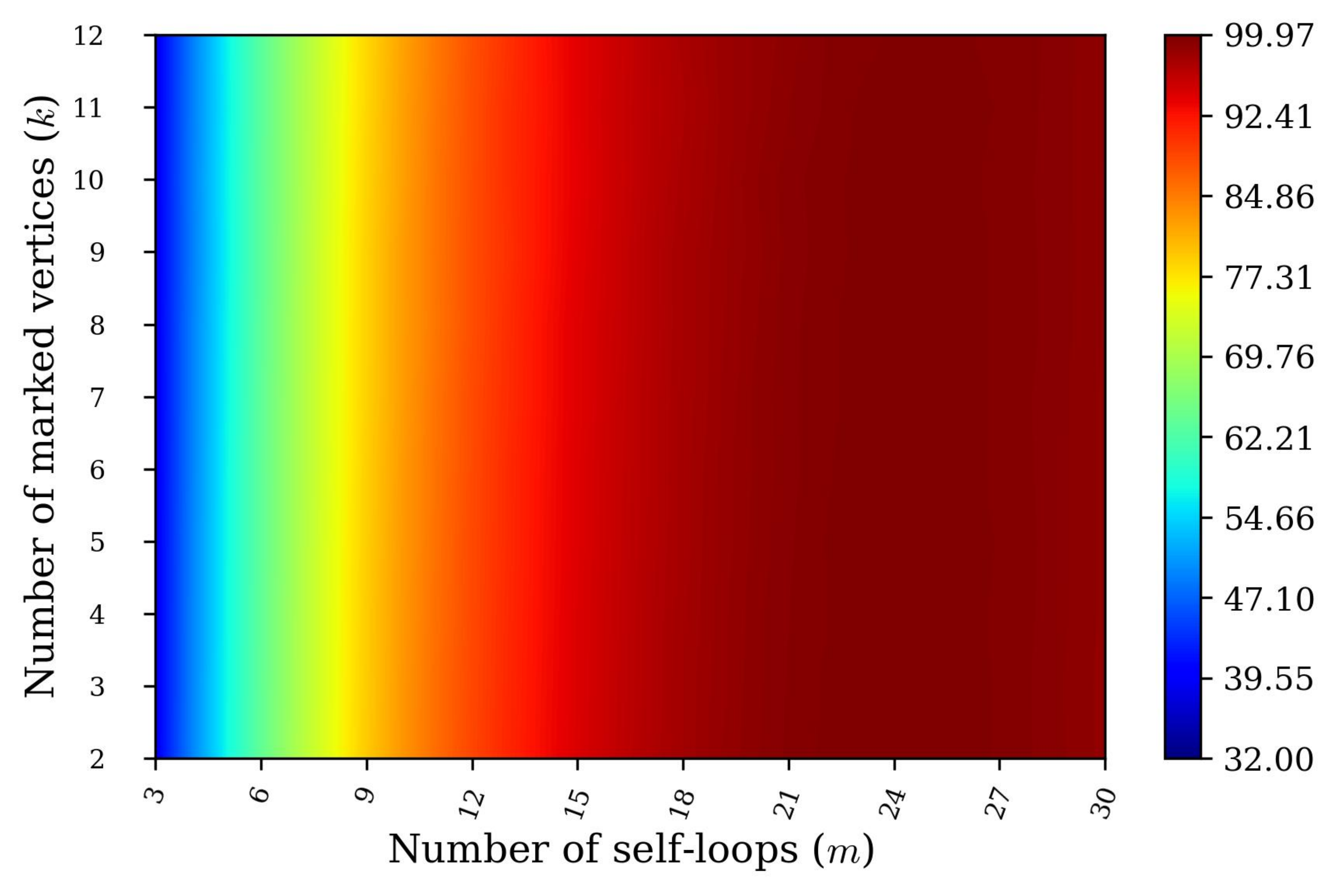}
    \caption{The success probability of MSLQW-PPI for weight value $l = (n^{2}/N)\cdot k$ for $s = 2$ inverted self-loops and $3 \leqslant m \leqslant 30$.}
    \label{fig:probability-distribution-non-neighbors-n-2-N-k-s-2}
\end{figure*}

\begin{figure*}[]
\centering
\subfloat[$s = 3, m = 36$]{\includegraphics[width=8cm]{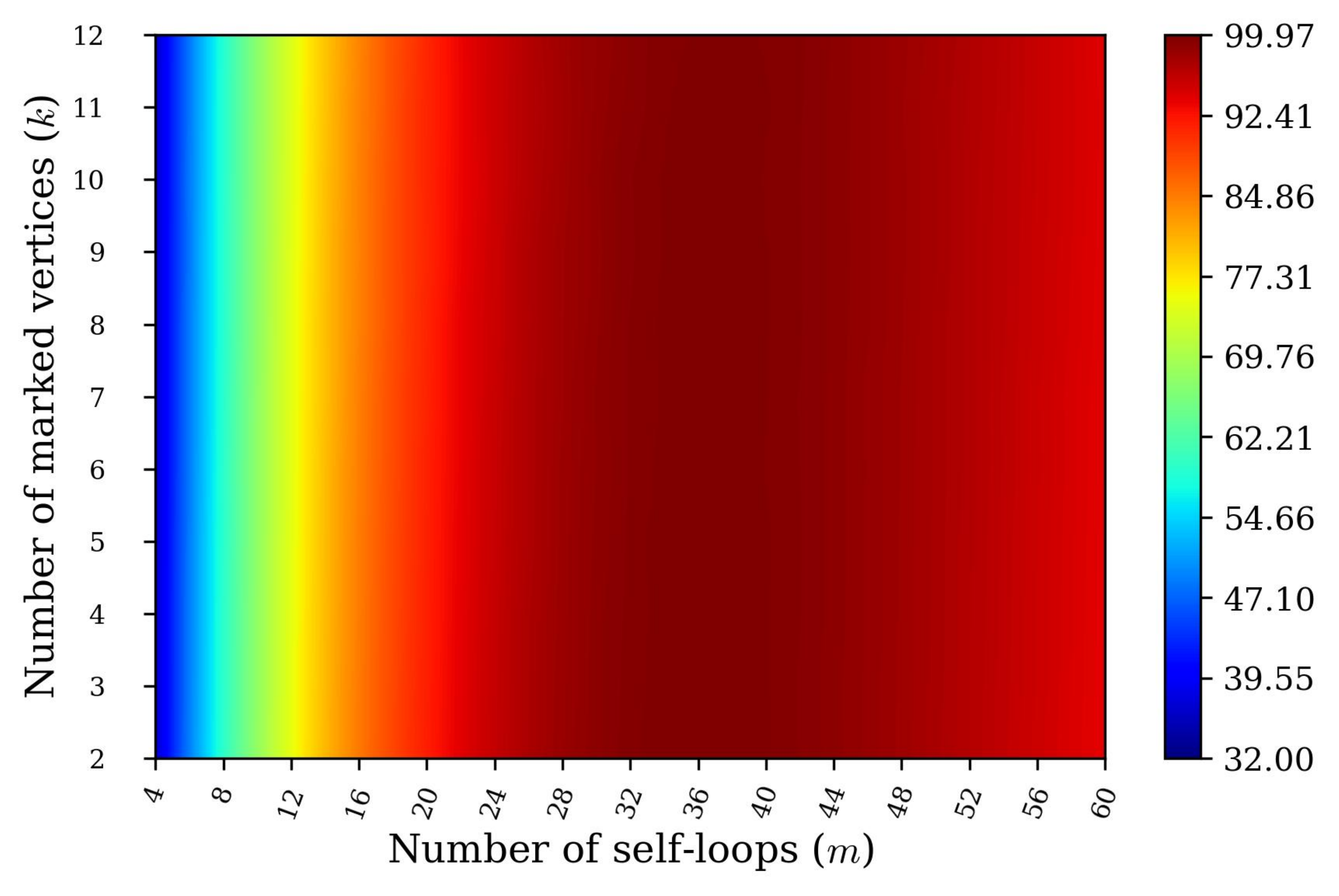}
\label{fig:probability-distribution-non-neighbors-n-2-N-k-s-3}}
\subfloat[$s = 4, m = 48$]{\includegraphics[width=8cm]{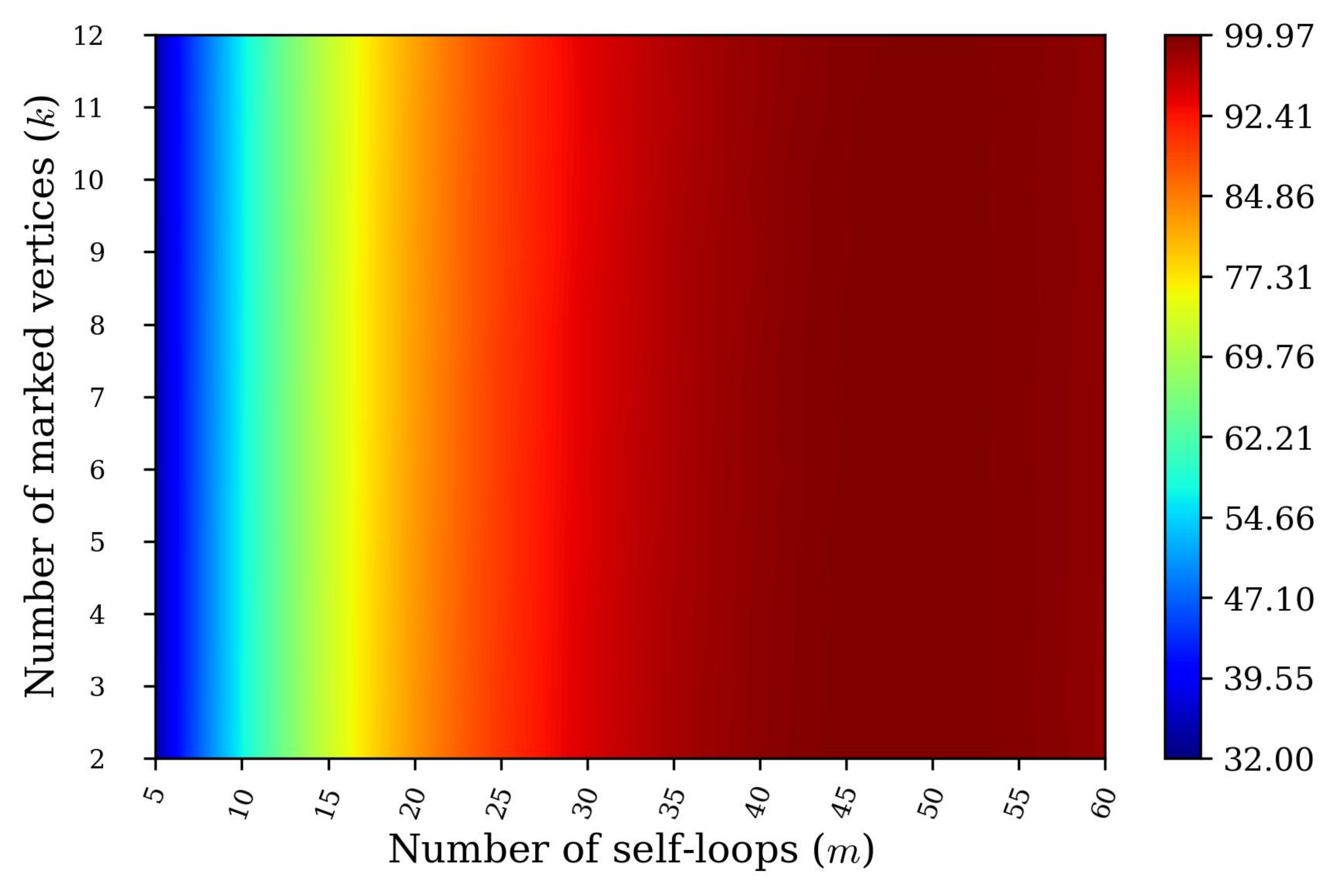}
\label{fig:probability-distribution-non-neighbors-n-2-N-k-s-4}}
\\
\subfloat[$s = 5, m = 60$]{\includegraphics[width=8cm]{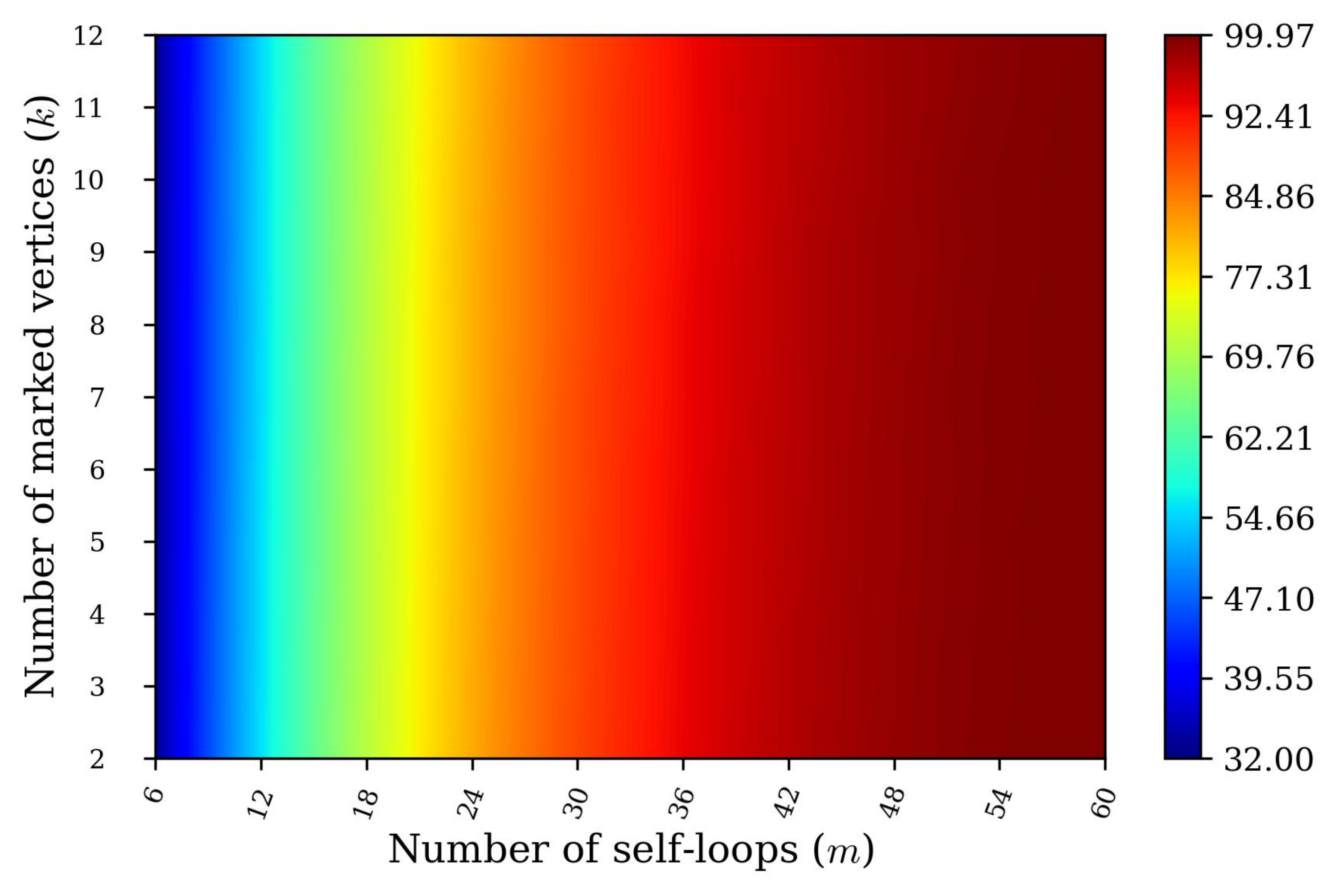}
\label{fig:probability-distribution-non-neighbors-n-2-N-k-s-5}}
\caption{The success probability of MSLQW-PPI for weight value $l = (n^{2}/N)\cdot k$. In Figures \protect\subref{fig:probability-distribution-non-neighbors-n-2-N-k-s-3}, \protect\subref{fig:probability-distribution-non-neighbors-n-2-N-k-s-4} and \protect\subref{fig:probability-distribution-non-neighbors-n-2-N-k-s-5} $s = 3,4,5$  and $4 \leqslant m \leqslant 60$, respectively.}
\label{fig:probability-distribution-non-neighbors-n-2-N-k-s-2-3-4-5}
\end{figure*}

The quantity of $m$ is calculated as follows: $m = s \cdot n$, where $s$ is the number of self-loops with its phases inverted, and $n$ is the degree of the hypercube. New experiments were performed to confirm this hypothesis. Fig.~\ref{fig:probability-distribution-non-neighbors-n-2-N-k-s-2} shows the result of MSLQW-PPI for $2 \leqslant s \leqslant 5$, and Fig.~\ref{fig:probability-distribution-non-neighbors-n-2-N-k-s-2-3-4-5} shows the result of MSLQW-PPI for $s = 2, 3, 4, 5$. For the number of inverted self-loops $s = 2, 3, 4, 5$ and $n = 12$, $m = 24, 36, 48, 60$ self-loops were needed to achieve maximum success probabilities.

\begin{table*}[]
\centering
\caption{Comparison between probabilities of success for searching non-adjacent marked vertices and the number $m$ of self-loops. The results presented here refer to the search using partial phase inversion of the target state. From $k = 4$ marked vertices, the success probabilities for weight $l = n/N$ remain below $p = 0.66$, while for the other three weight values, the probability of success stays above $p = 0.99$. The acronym cv means the coefficient of variation.}
\label{tab:comparison-weights}
\begin{tabular}{lcccccccccccc}
\toprule
 & \multicolumn{12}{c}{Self-loop weights} \\ \cmidrule{2-13} 
 & \multicolumn{3}{c}{$l = n/N$} & \multicolumn{3}{c}{$l = (n/N)\cdot k$} & \multicolumn{3}{c}{$l = n^{2}/N$} & \multicolumn{3}{c}{$l = (n^{2}/N)\cdot k$} \\ \cmidrule(lr){2-4}\cmidrule(lr){5-7}\cmidrule(lr){8-10}\cmidrule(lr){11-13} 
k & p     & $m$ &  cv              & p     & $m$ & cv  & p     & $m$ & cv           & p     & $m$  & cv \\ \midrule
2 & 0.887 & 1 &  1.643e-04  & 0.999 & 1 &  8.259e-05  & 0.999 & 6 & 5.234e-05 & 0.999 & 12 & 2.810e-04 \\
3 & 0.750 & 1 &  7.037e-04  & 0.999 & 1 &  1.144e-04  & 0.999 & 4 & 1.050e-04 & 0.999 & 12 & 2.095e-04 \\
4 & 0.663 & 1 &  3.882e-03  & 0.999 & 1 &  1.221e-04  & 0.998 & 3 & 1.922e-04 & 0.999 & 12 & 6.324e-04 \\ \bottomrule
\end{tabular}
\end{table*}

\begin{figure*}[]
\centering
\subfloat[One self-lop.]{\includegraphics[width=8.5cm]{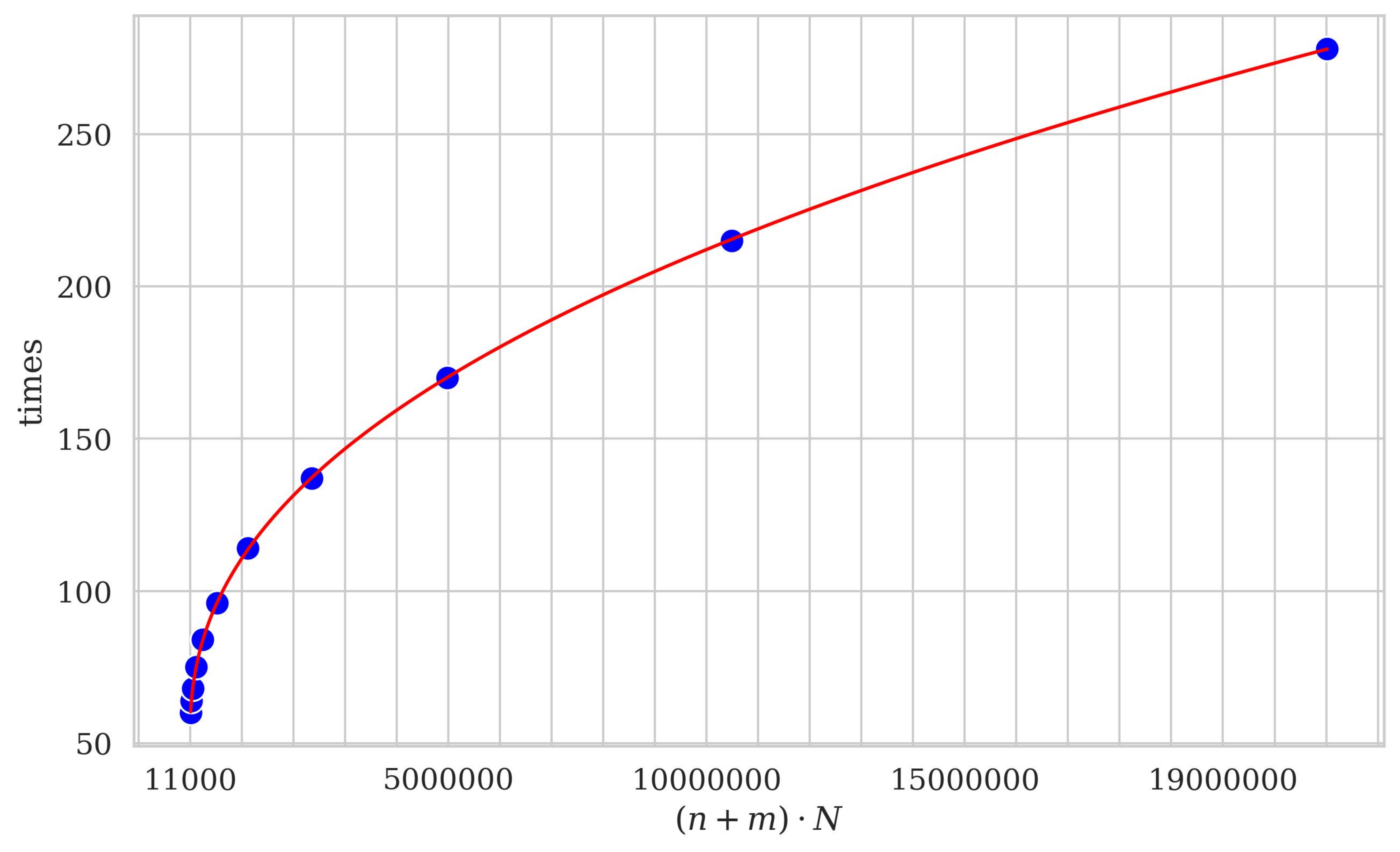}
\label{fig:hypercube-curve-fitting-one-self-loops-non-neighbors-n-2-N-k}}
\subfloat[Multiple self-loops.]{\includegraphics[width=8.5cm]{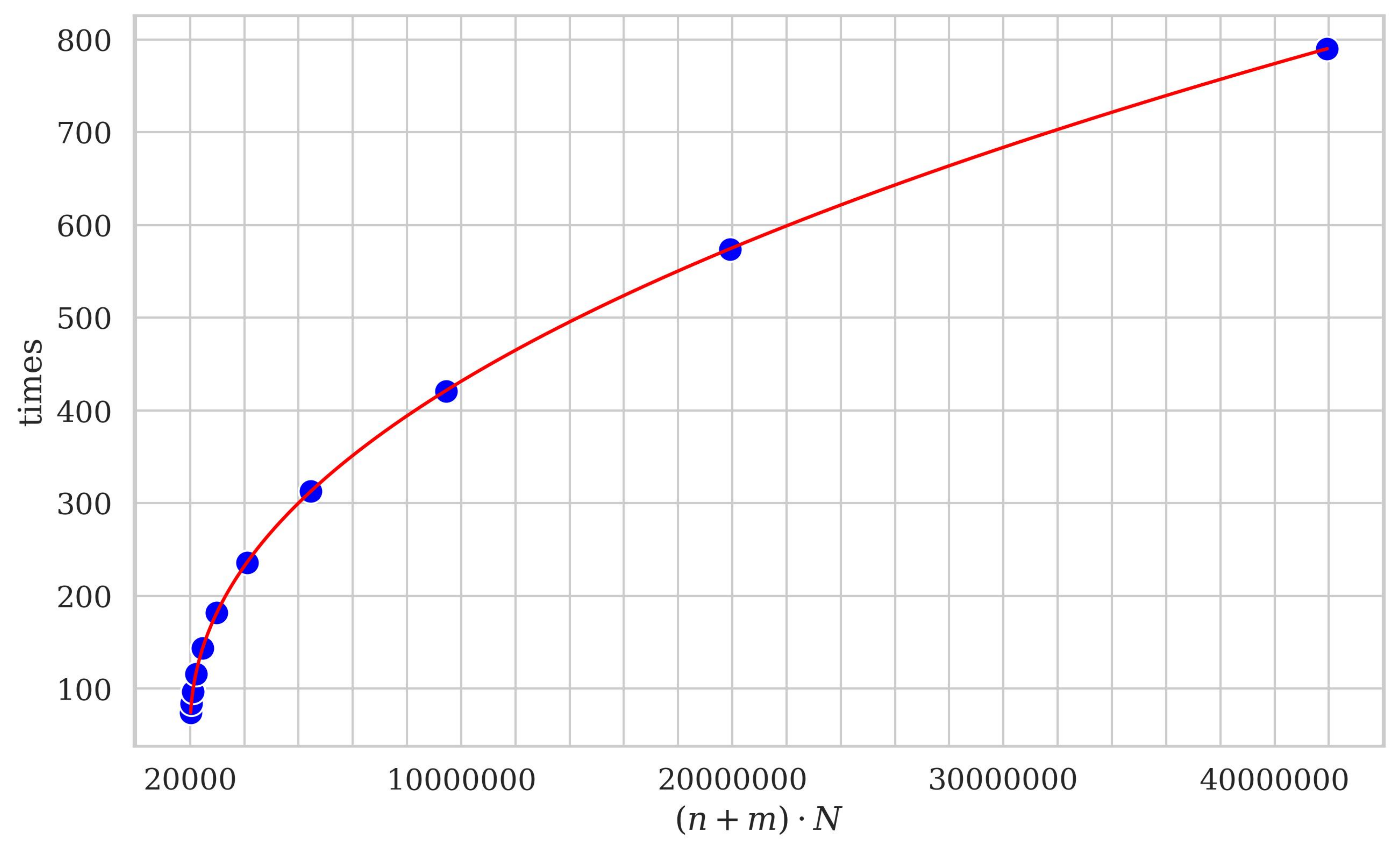}
\label{fig:hypercube-curve-fitting-multiple-self-loops-non-neighbors-n-2-N-k}}
\caption{The time complexity of the algorithm relative to the size of the hypercube. The solid red line represents the estimated curve and the blue dots are the numerical simulation values of the quantum walk.}
\label{fig:hypercube-curve-fitting-one-multiple-self-loops-non-neighbors-n-2-N-k}
\end{figure*}

To obtain the complexity of the proposed algorithm, two analyses were performed. The first analysis is about knowing how the runtime complexity behaves as $N = 2^{n}$ is changed. The second analysis is about knowing how the runtime complexity behaves as $m$ self-loops are added at each vertex of the hypercube. In both analyses, the weight $l = (n^{2}/N)\cdot k$ was used. Fig. \ref{fig:hypercube-curve-fitting-one-multiple-self-loops-non-neighbors-n-2-N-k} shows the results of the quantum walk applied to eleven hypercubes of degrees $n = \{10,...20\}$ respectively. Fig. \ref{fig:hypercube-curve-fitting-one-self-loops-non-neighbors-n-2-N-k} presents the results obtained using only one self-loop, and Fig. \ref{fig:hypercube-curve-fitting-multiple-self-loops-non-neighbors-n-2-N-k} shows the results obtained with the MSLQW-PPI.

Considering the use of a single self-loop at each vertex of the hypercube we have a cost of square root. The adjustments show that the running time $t$ is
\begin{align*}
    t = c_{1} \cdot \sqrt{((n + \text{m}) \cdot N) ^{c_{2}}} + c_{3},
\end{align*}
where $c_{1}, c_{2} \text{ and } c_{3} \in \mathbb{R}$. Then, numerical simulations suggest that the running time is $O(\sqrt{((n + m)\cdot N })$. This also occurs when using multiple self-loops and partial phase inversion. Maximum success probabilities stay close to $1$ as $n$ changes. Note that when we fix the number of self-loops the running time of the algorithm does not on the dimensions of the hypercube. The fit curves are

\begin{align*}
    t_{a} = 0.1408 \cdot \sqrt{((n + \text{m}) \cdot N)^{0.873}} + 52.0675
\end{align*}
and
\begin{align*}
    t_{b} = 0.2105 \cdot \sqrt{((n + \text{m}) \cdot \textrm{N})^{0.9299}} + 53.0933.
\end{align*}
Where $t_{a}$ is the cost for simulations with only one self-loop -- Fig. \ref{fig:hypercube-curve-fitting-one-self-loops-non-neighbors-n-2-N-k}, and $t_{b}$ is the cost of simulation with the best self-loop quantity (between 1 to 30) for each hypercube size -- Fig. \ref{fig:hypercube-curve-fitting-multiple-self-loops-non-neighbors-n-2-N-k}.

\begin{figure*}[]
    \centering
    \subfloat[]{\includegraphics[width=5.5cm]{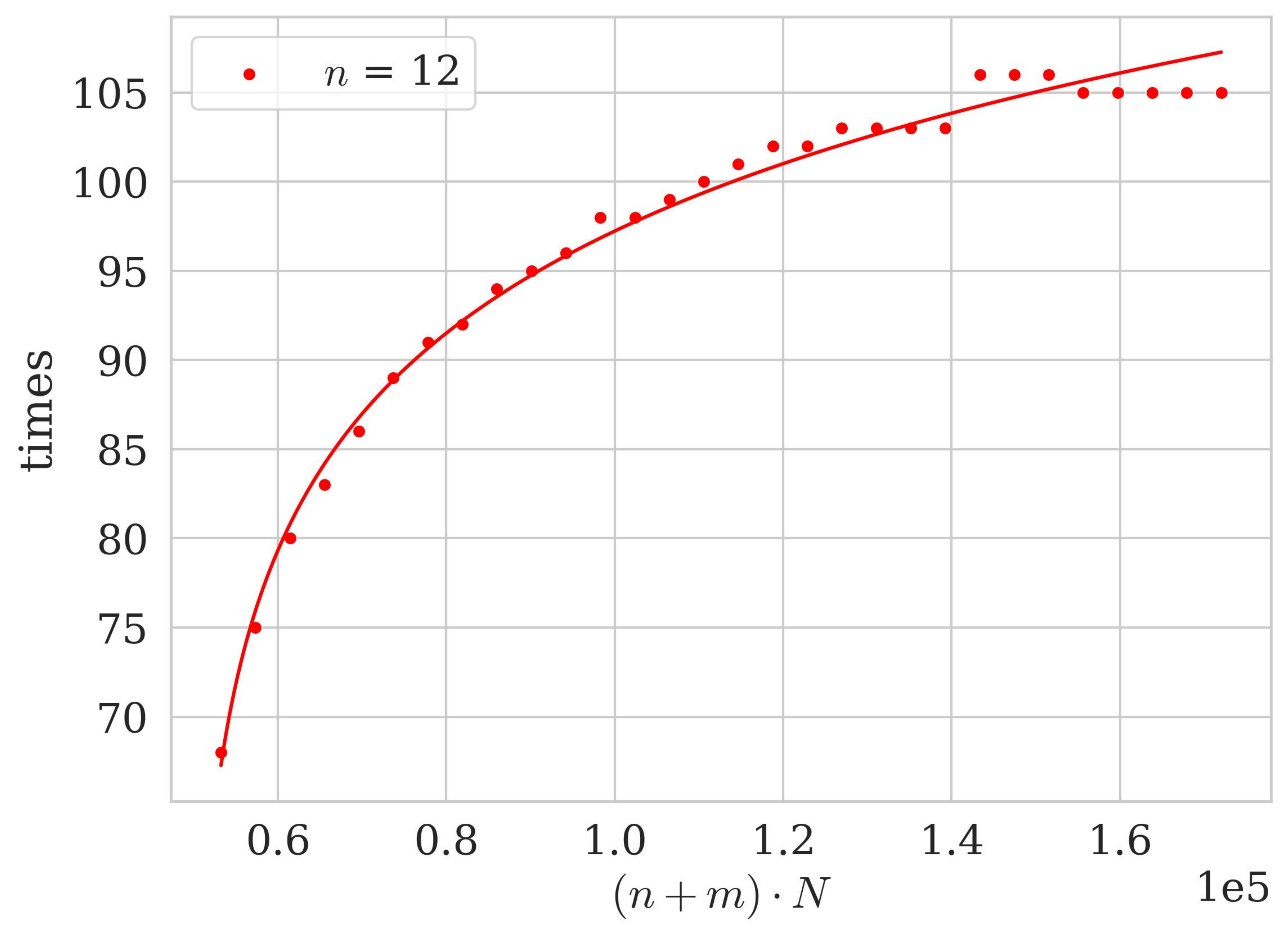}
    \label{fig:hypercube_curve_fitting_self-loops_n_12}}
    \subfloat[]{\includegraphics[width=5.5cm]{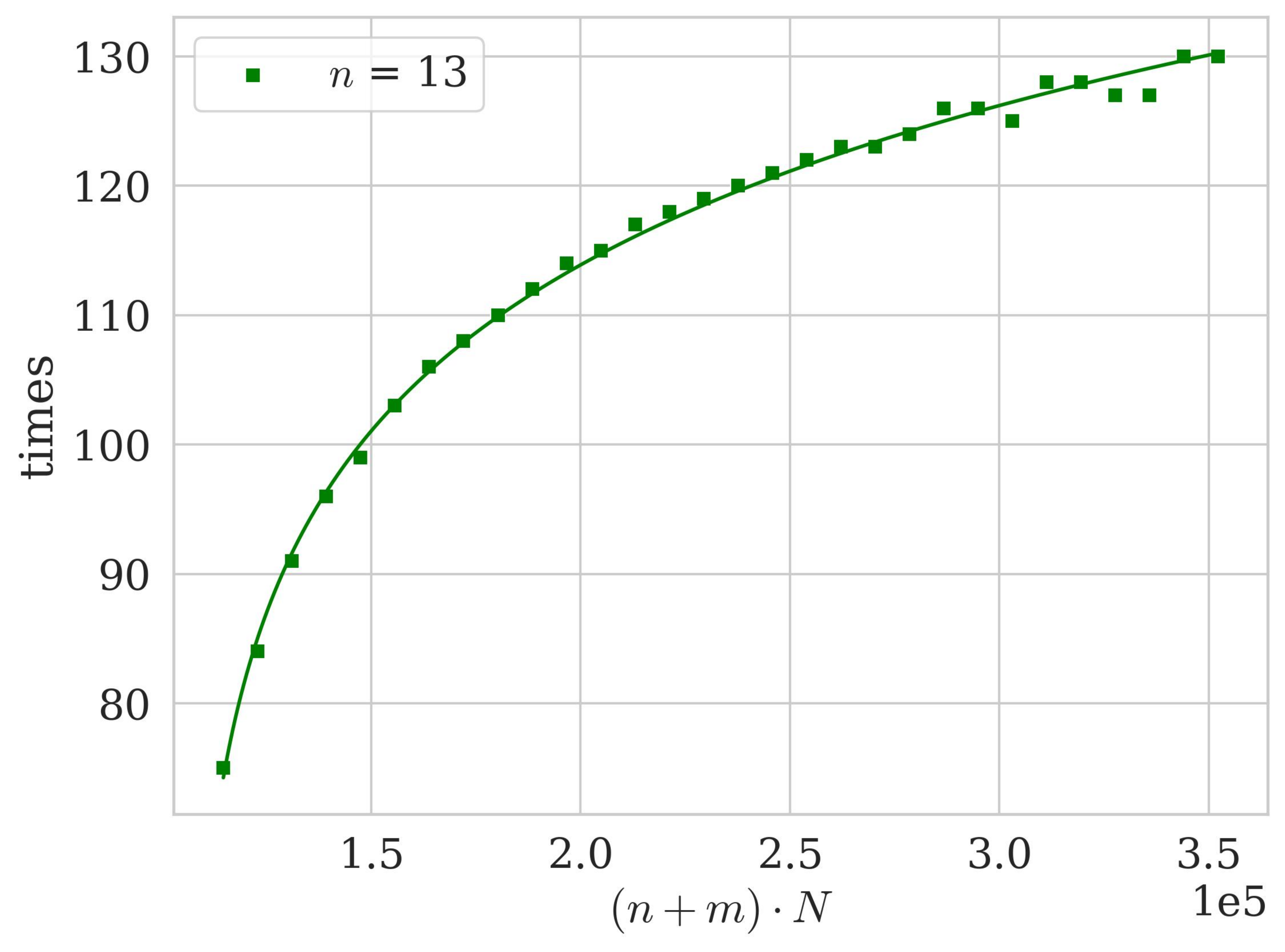}
    \label{hypercube_curve_fitting_self-loops_n_13}}
    \subfloat[]{\includegraphics[width=5.5cm]{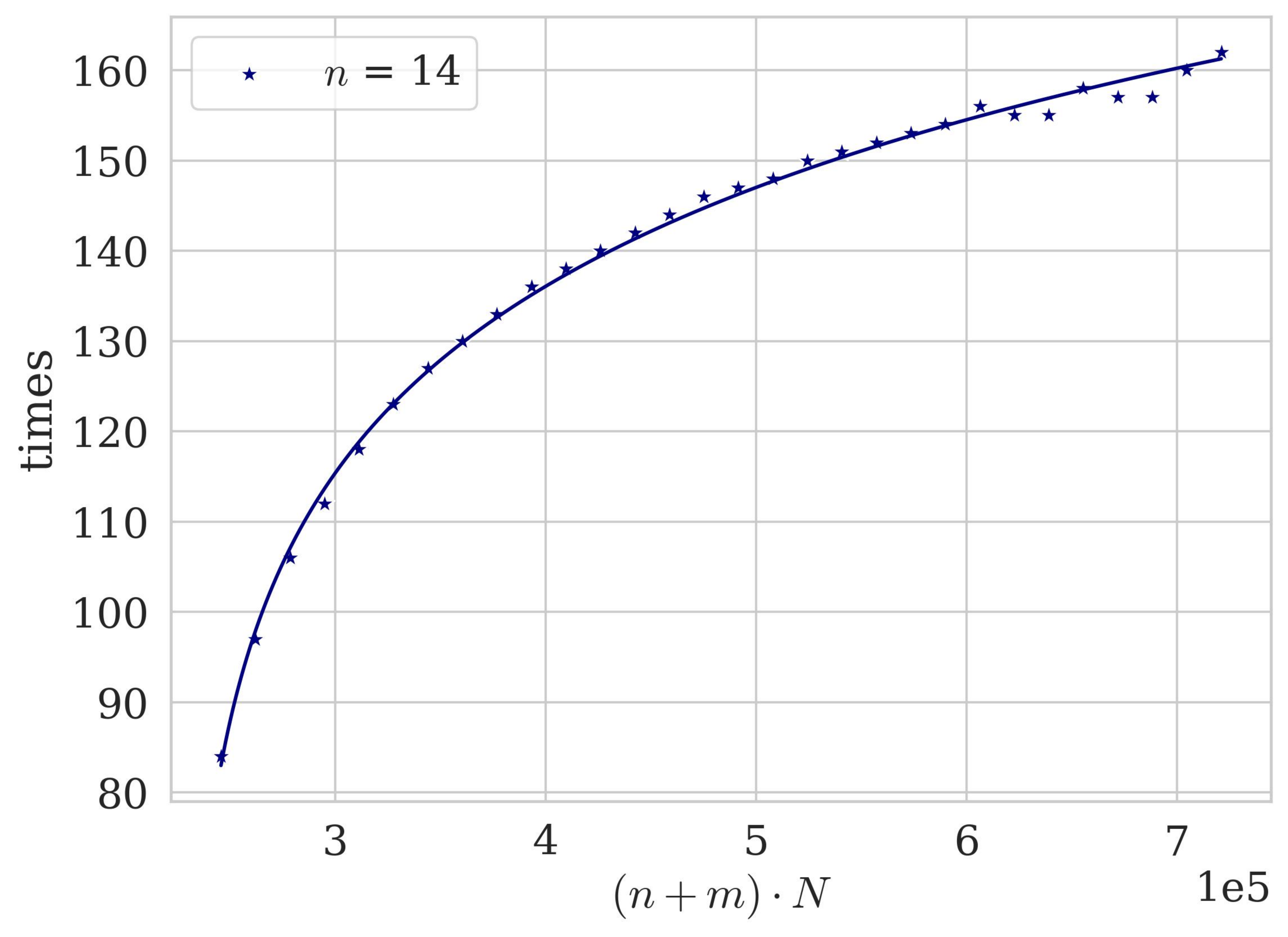}
    \label{fig:hypercube_curve_fitting_self-loops_n_14}}\\
    \subfloat[]{\includegraphics[width=5.5cm]{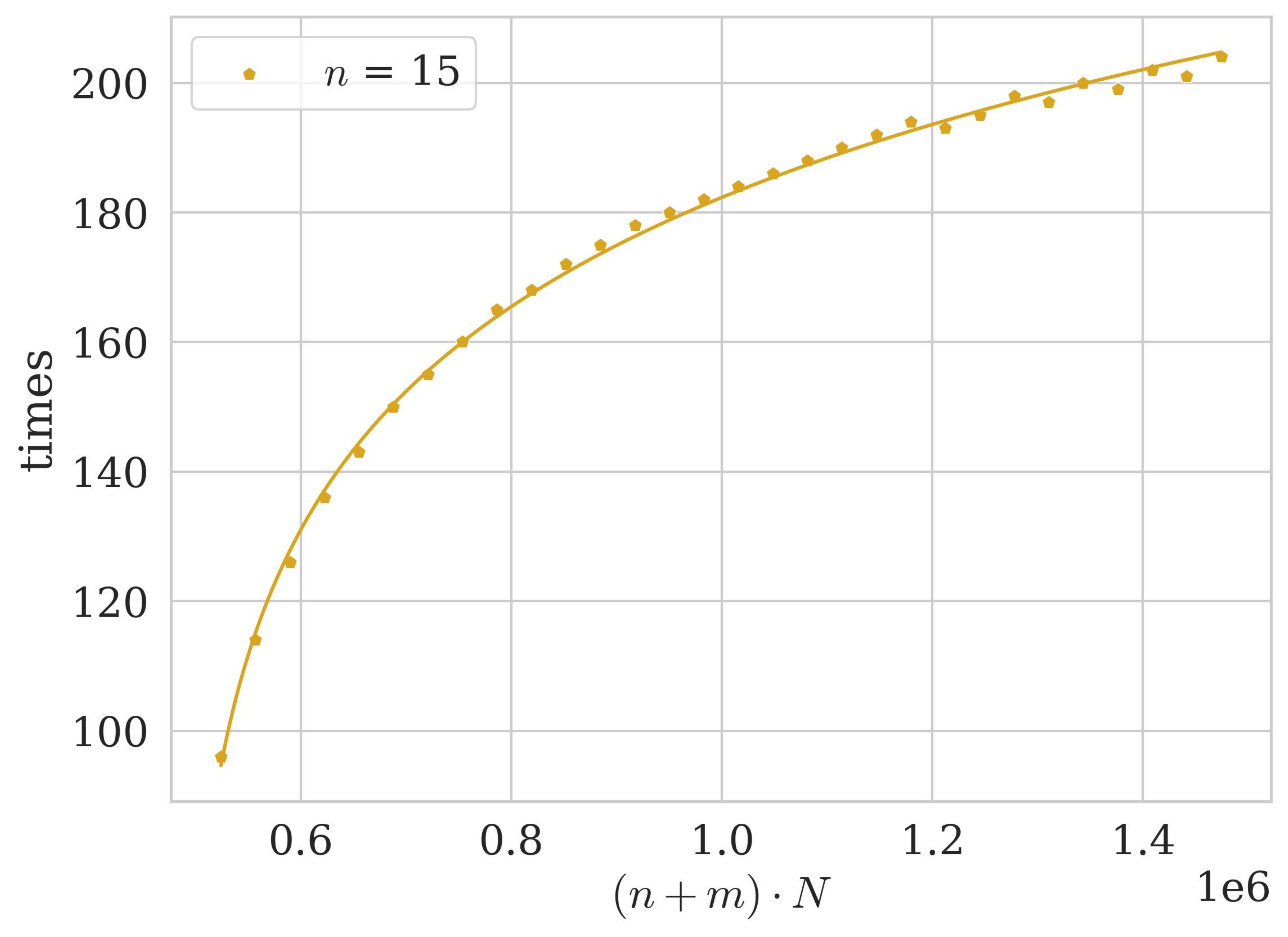}
    \label{fig:hypercube_curve_fitting_self-loops_n_15}}
    \subfloat[]{\includegraphics[width=5.5cm]{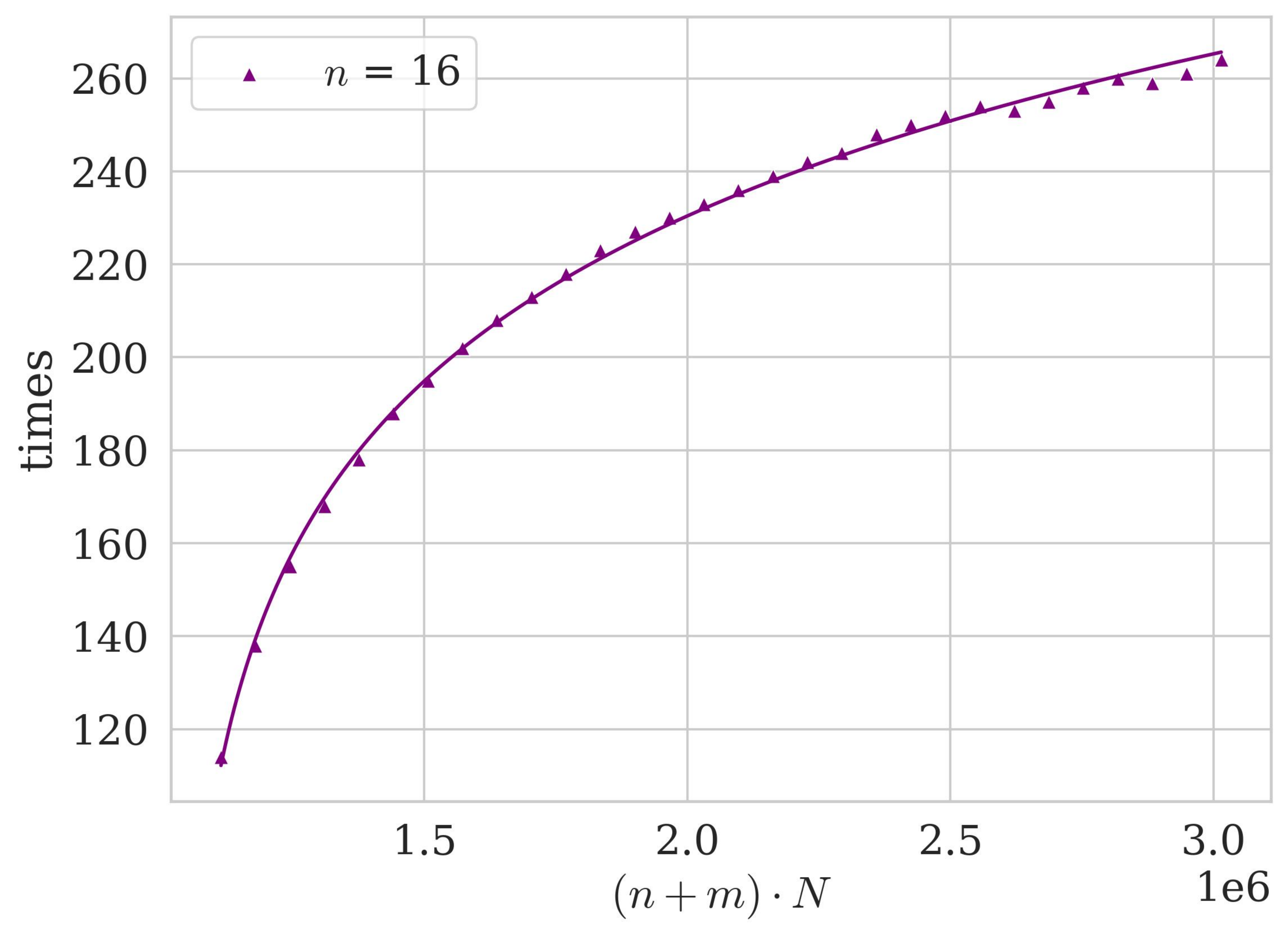}
    \label{fig:hypercube_curve_fitting_self-loops_n_16}}
    \caption{The algorithm's time complexity concerning the number of self-loops at each vertex. The solid lines represent the estimated curves. The points are the values from numerically simulating the quantum walk. For each figure, $n$ is a constant.}
    \label{fig:hypercube-curve-fitting-self-loops-non-neighbors-n-2-N-k}
\end{figure*}

Once the number of vertices of the hypercube is defined, the complexity is logarithmic as we can see. Fig. \ref{fig:hypercube-curve-fitting-self-loops-non-neighbors-n-2-N-k} shows the results of the MSLQW-PPI applied to five hypercubes of degrees $n = \{12,...16\}$ respectively. On each of these hypercubes, thirty quantum walks were performed. The results show how many times the evolution operator needs to be applied to obtain the maximum probability of success concerning the number of self-loops for each of the hypercube sizes. The adjustments show that the running time $t$ is
\begin{equation*}
    t = c_{1} \cdot \log{((n+m)\cdot N + c_{2})} + c_{3},
\end{equation*}
where $c_{1}, c_{2} \text{ and } c_{3} \in \mathbb{R}$. Then, numerical simulations suggest that the running time is $O(\log{((n+\text{m})\cdot N}))$, where $n$ is a constant representing the dimension of the hypercube in use, $m$ is the number of self-loops, and $N$ is also a constant representing the total number of vertices of the hypercube. The run times are shown in Table \ref{tab:log-fittings}. However, if all constants are discarded, the computational cost for this situation can be approximated to $O(\log{(m)})$.

\begin{table}[]
\centering
\caption{Algorithm complexity adjustments, $m$ is the number of self-loops.}
\label{tab:log-fittings}
\begin{tabular}{ccrr}
\toprule
\multicolumn{4}{c}{$c_{1} \cdot \log{((n+m)\cdot N + c_{2})} + c_{3}$} \\ \hline
n & \multicolumn{1}{c}{$c_{1}$} & \multicolumn{1}{c}{$c_{2}$} & \multicolumn{1}{c}{$c_{3}$} \\ \midrule
12 & 11.6237 & -49296.2873 & -28.8548 \\
13 & 17.0371 & -105453.4444  & -81.3999 \\
14 & 23.5898 & -227612.0627 & -148.5875 \\
15 & 34.2267 & -484336.2147 & -268.0318 \\
16 & 49.4194 & -1024918.6020 & -451.1252 \\ \bottomrule
\end{tabular}
\end{table}

\section{Conclusions}
\label{sec:conclusions}

The lackadaisical quantum walk has a strong dependence on the self-loop weight value. We can see this fact in the results presented in all figures. For the cases where the phases of all self-loops are flipped, using a single self-loop per vertex is better. We can affirm the same for the results shown in Fig.~\ref{fig:probability-distribution-non-neighbors-all-signs-weights-n-N-and-n-N-k}.

When \citet{wong2015grover} proposed the lackadaisical quantum walk, he included $m$ integer self-loops at each vertex in the complete graph, which according to Rhodes and \citet{rhodes2020search} is equivalent to having $m$ unweighted self-loops. However, \citet{wong2017coined} redefined the lackadaisical quantum walk, where each vertex with $m$ integer self-loops can be reduced to a quantum walk where each vertex has a single self-loop of real weight $l$. Note that the number of self-loops and the weight value $l$ are the main characteristics of this quantum walk. Our strategy differs in that we apply partial phase inversion and consider a real-valued weight $l$ and equally distribute it in $m$ self-loops in each vertex, \textit{i.e.}, $m$ self-loops of non-integer weight value. In addition to using the two existing weight values in the literature to perform our experiments, we can contribute to the proposal of two new weight values. As the vertex degree is one of the parameters used in many works to define the weights assigned to self-loops, we suggest a modification that adds an exponent $2$ to the numerator.

Quantum interference is essential in developing quantum algorithms \citep{mcmahon2007quantum}. The interference caused by the phase inversion operation, jointly with the average inversion operation, amplifies the target states' amplitudes. It allows Grover's search algorithm a certain probability of finding the desired state \citep{grover1996fast}. The self-loops are redundant elements within the structure that compose the states of the quantum system. 

When we interfered jointly by flipping the phase of the $m$ self-loops, making them indistinguishable, we saw that it was equivalent to using a single self-loop. However, when we invert the phase of one of the $m$ self-loops, we empirically observe that the behavior caused by the constructive and destructive interference between the $m$ self-loops are analogous to those observed between the states that represent the vertices as a whole, \textit{i.e.}, most of the energy of the $m$ self-loops is retained in the phase-inverted self-loop. The results indicate that the phase inversion of a single self-loop $\circlearrowleft_{\tau=j}$ is sufficient to obtain the results presented in this work. However, it is possible to find maximum success probabilities close to $1$ by inverting $1 < s < m$ self-loops and using a total of $\text{m} = s * n$ self-loops.

To obtain the average behavior based on the relative position of the marked vertices, each simulation was performed on a sample that contains a number $k$ of distinctly marked vertices, that is, without replacement. Thus, verifying whether the relative position of non-adjacent marked vertices influences the results is possible. According to the results, the relative position of non-adjacent marked vertices does not significantly affect considering a numerical precision of four digits. Furthermore, the coefficient of variation indicates that the number $m$ of self-loops and the number $k$ of marked vertices influence the results. We used the coefficient of variation to analyze the results' dispersion level. We observe that for MSLQW-PPI when we have the slightest standard deviations, the maximum probabilities of success scale to values close to $1$. The percentage variations around the mean are not significant. However, the behavior is stable for MSLQW-PPI, where there are slight variations. The results are more influenced by the number of marked vertices and the number of self-loops than by the relative position of the non-adjacent marked vertices. As we can see from the results, the multiple self-loops are an essential tool to improve the success probability of searching multiple marked vertices on the hypercube. Parameters such as weight value, weight distribution strategies, and phase inversion operation contributed to the results of this work.

In this work, the marked vertices are all non-adjacent. According to \citet{souza2021lackadaisical}, the type of marked vertices can interfere with the result of the quantum walk on the hypercube. Therefore, as a supplementary work, we have used the MSLQW-PPI to search for multiple adjacent marked vertices on the hypercube. In this sense, some preliminary results can be found in the pre-print \citep{de2023search}.  We also intend to apply this methodology to evaluate the MSLQW-PPI in other $d$-regular structures, for example, the Johnson graph~\citep{peng2024lackadaisical}. We intend to analyze other exponent values $\alpha$ for the composition of the weights $l = \{n^{\alpha}/N, (n^{\alpha}/N)\cdot k\}$, including real values, \textit{i.e.}, $\{\alpha \in \mathbb{R} \mid \alpha\neq 1\}$. It is possible to use an evolutionary search algorithm to define the best exponent $\alpha$ value capable of maximizing the probability of success and minimizing the number of self-loops influencing the current proposal. We intend to propose other strategies to distribute weight values, such as providing distinct self-loop weight values for marked vertices. Another possible path for deeper investigation is to consider the partial inversion of edges $i$ that are not self-loops. Furthermore, proposing a variation of Grover's search algorithm could also be a promising direction for future efforts.

\section*{Acknowledgments}
\label{sec:acknowledgments}
Acknowledgments to the Science and Technology Support Foundation of Pernambuco (FACEPE)- Brazil, The Brazilian National Council for Scientific and Technological Development (CNPq), and the Coordena\c{c}\~{a}o de Aperfei\c{c}oamento de Pessoal de N\'{i}vel Superior - Brasil (CAPES) - Finance Code 001 by their financial support to the development of this research.

\bibliographystyle{unsrtnat}
\bibliography{references}  %%% Uncomment this line and comment out the ``thebibliography'' section below to use the external .bib file (using bibtex) .

%%% Uncomment this section and comment out the \bibliography{references} line above to use inline references.
% \begin{thebibliography}{1}

% 	\bibitem{kour2014real}
% 	George Kour and Raid Saabne.
% 	\newblock Real-time segmentation of on-line handwritten arabic script.
% 	\newblock In {\em Frontiers in Handwriting Recognition (ICFHR), 2014 14th
% 			International Conference on}, pages 417--422. IEEE, 2014.

% 	\bibitem{kour2014fast}
% 	George Kour and Raid Saabne.
% 	\newblock Fast classification of handwritten on-line arabic characters.
% 	\newblock In {\em Soft Computing and Pattern Recognition (SoCPaR), 2014 6th
% 			International Conference of}, pages 312--318. IEEE, 2014.

% 	\bibitem{hadash2018estimate}
% 	Guy Hadash, Einat Kermany, Boaz Carmeli, Ofer Lavi, George Kour, and Alon
% 	Jacovi.
% 	\newblock Estimate and replace: A novel approach to integrating deep neural
% 	networks with existing applications.
% 	\newblock {\em arXiv preprint arXiv:1804.09028}, 2018.

% \end{thebibliography}

% TODO: Retomar daqui. Corrigir braket <*>

\appendix
\section{Evaluation of the Oracle application in three possible scenarios}
\label{sec:appendix-oracle-evaluation}

Let us evaluate the application of the oracle in the three possible scenarios with respect to the edges. For each scenario, we also apply the oracle to unmarked vertices to show its general effectiveness.
\vspace{-0.2cm}

\begin{scenario}
\label{sc:scenario-one}
The first scenario shows the oracle applied in states $\ket{\circlearrowleft_{j},\Vec{x}}$ which represents the self-loop to be inverted.

\noindent$\bullet$ $\omega = \Vec{x}$, $\epsilon \neq \circlearrowleft_{j}$, and $\{\circlearrowleft_{\tau} = \circlearrowleft_{j}\}$.

\begin{align}
\label{eq:scenario-one-self-loop-target-invertion}
    \begin{split}
        Q\sum_{j=0}^{s-1}\ket{\circlearrowleft_{j},\Vec{x}} &= (I_{(n+\text{m})\cdot N} - 2\ket{\epsilon,\omega}\bra{\epsilon,\omega} -2\ket{\circlearrowleft_{\tau=j},\omega}\bra{\circlearrowleft_{\tau=j},\omega}) \sum_{j=0}^{s-1} \ket{\circlearrowleft_{j},\Vec{x}}\\
        &= \sum_{j=0}^{s-1}\ket{\circlearrowleft_{j},\Vec{x}} - 2\ket{\epsilon,\omega}\bra{\epsilon,\omega}\sum_{j=0}^{s-1}\ket{\circlearrowleft_{j},\Vec{x}} -2\ket{\circlearrowleft_{\tau=j},\omega}\bra{\circlearrowleft_{\tau=j},\omega}\sum_{j=0}^{s-1}\ket{\circlearrowleft_{j},\Vec{x}} \\
        &= \sum_{j=0}^{s-1}\ket{\circlearrowleft_{j},\Vec{x}} - 2\ket{\epsilon,\omega}\cdot 0 -2\ket{\circlearrowleft_{\tau=j},\omega} \cdot 1 \\
        &= -\sum_{j=0}^{s-1}\ket{\circlearrowleft_{j},\Vec{x}}
    \end{split}
\end{align}

\noindent$\bullet$ $\omega \neq \Vec{x}$, $\epsilon \neq \circlearrowleft_{j}$, and $\{\circlearrowleft_{\tau} = \circlearrowleft_{j}\}$.

\begin{align}
\label{eq:scenario-one-self-loop-target-non-invertion}
    \begin{split}
        Q\sum_{j=0}^{s-1}\ket{\circlearrowleft_{j},\Vec{x}} &= \sum_{j=0}^{s-1}\ket{\circlearrowleft_{j},\Vec{x}} - 2\ket{\epsilon,\omega}\bra{\epsilon,\omega}\sum_{j=0}^{s-1}\ket{\circlearrowleft_{j},\Vec{x}}-2\ket{\circlearrowleft_{\tau=j},\omega}\bra{\circlearrowleft_{\tau=j},\omega}\sum_{j=0}^{s-1}\ket{\circlearrowleft_{j},\Vec{x}} \\
        &= \sum_{j=0}^{s-1}\ket{\circlearrowleft_{j},\Vec{x}} - 2\ket{\epsilon,\omega}\cdot 0 -2\ket{\circlearrowleft_{\tau=j},\omega} \cdot 0 \\
        &= \sum_{j=0}^{s-1}\ket{\circlearrowleft_{j},\Vec{x}}
    \end{split}
\end{align}

\end{scenario}

\begin{scenario}
\label{sc:scenario-two}

The second scenario shows the application of the oracle in the $j$th self-loop $\ket{\circlearrowleft_{j},\Vec{x}}$, where $\tau \neq j$.

\noindent$\bullet$ $\omega = \Vec{x}$, $\epsilon \neq \circlearrowleft_{j}$, and $\{\circlearrowleft_{\tau} \neq \circlearrowleft_{j}\}$.

\begin{align}
\label{eq:scenario-two-jth-self-loop-omega-neq-x}
    \begin{split}
        Q\sum_{j=s}^{m-1}\ket{\circlearrowleft_{j},\Vec{x}} &= (I_{(n+\text{m})\cdot N} - 2\ket{\epsilon,\omega}\bra{\epsilon,\omega}-2\ket{\circlearrowleft_{\tau=j},\omega}\bra{\circlearrowleft_{\tau=j},\omega}) \sum_{j=s}^{m-1}\ket{\circlearrowleft_{j},\Vec{x}}\\
        &= \sum_{j=s}^{m-1}\ket{\circlearrowleft_{j},\Vec{x}} - 2\ket{\epsilon,\omega}\bra{\epsilon,\omega}\sum_{j=s}^{m-1}\ket{\circlearrowleft_{j},\Vec{x}}-2\ket{\circlearrowleft_{\tau=j},\omega}\bra{\circlearrowleft_{\tau=j},\omega}\sum_{j=s}^{m-1}\ket{\circlearrowleft_{j},\Vec{x}} \\
        &= \sum_{j=s}^{m-1}\ket{\circlearrowleft_{j},\Vec{x}} - 2\ket{\epsilon,\omega}\cdot 0 -2\ket{\circlearrowleft_{\tau=j},\omega} \cdot 0 \\
        &= \sum_{j=s}^{m-1}\ket{\circlearrowleft_{j},\Vec{x}}
    \end{split}
\end{align}

\noindent$\bullet$ $\omega \neq \Vec{x}$, $\epsilon \neq \circlearrowleft_{j}$, and $\{\circlearrowleft_{\tau} \neq \circlearrowleft_{j}\}$.

\begin{align}
\label{eq:scenario-two-jth-self-loop-omega-eq-x}
    \begin{split}
        Q\sum_{j=s}^{m-1}\ket{\circlearrowleft_{j},\Vec{x}} &= \sum_{j=s}^{m-1}\ket{\circlearrowleft_{j},\Vec{x}} - 2\ket{\epsilon,\omega}\bra{\epsilon,\omega}\sum_{j=s}^{m-1}\ket{\circlearrowleft_{j},\Vec{x}}-2\ket{\circlearrowleft_{\tau=j},\omega}\bra{\circlearrowleft_{\tau=j},\omega}\sum_{j=s}^{m-1}\ket{\circlearrowleft_{j},\Vec{x}} \\
        &= \sum_{j=s}^{m-1}\ket{\circlearrowleft_{j},\Vec{x}} - 2\ket{\epsilon,\omega}\cdot 0 -2\ket{\circlearrowleft_{\tau=j},\omega} \cdot 0 \\
        &= \sum_{j=s}^{m-1}\ket{\circlearrowleft_{j},\Vec{x}}
    \end{split}
\end{align}

\end{scenario}

\begin{scenario}
\label{sc:scenario-three}

The third scenario shows the application of the oracle on the $i$th non-loop edge $\ket{i,\Vec{x}}$, \textit{i.e.}, which is not a self-loop.

\noindent$\bullet$ $\omega = \Vec{x}$, $\epsilon = i$, and $\{\circlearrowleft_{\tau} \neq i\}$.

\begin{align}
\label{eq:scenario-three-ith-self-loop-omega-neq-x}
    \begin{split}
        Q\ket{i,\Vec{x}} &= (I_{(n+\text{m})\cdot N} - 2\ket{\epsilon,\omega}\bra{\epsilon,\omega}-2\ket{\circlearrowleft_{\tau=j},\omega}\bra{\circlearrowleft_{\tau=j},\omega}) \ket{i,\Vec{x}}\\
        &= \ket{i,\Vec{x}} - 2\ket{\epsilon,\omega}\braket{\epsilon,\omega|i,\Vec{x}}-2\ket{\circlearrowleft_{\tau=j},\omega}\braket{\circlearrowleft_{\tau=j},\omega|i,\Vec{x}} \\
        &= \ket{i,\Vec{x}} - 2\ket{\epsilon,\omega}\cdot 1 -2\ket{\circlearrowleft_{\tau=j},\omega} \cdot 0 \\
        &= -\ket{i,\Vec{x}}
    \end{split}
\end{align}

\noindent$\bullet$ $\omega \neq \Vec{x}$, $\epsilon = i$, and $\{\circlearrowleft_{\tau} \neq i\}$.

\begin{align}
\label{eq:scenario-three-ith-self-loop-omega-eq-x}
    \begin{split}
        Q\ket{i,\Vec{x}} &= \ket{i,\Vec{x}} - 2\ket{\epsilon,\omega}\braket{\epsilon,\omega|i,\Vec{x}}-2\ket{\circlearrowleft_{\tau=j},\omega}\braket{\circlearrowleft_{\tau=j},\omega|i,\Vec{x}} \\
        &= \ket{i,\Vec{x}} - 2\ket{\epsilon,\omega}\cdot 0 -2\ket{\circlearrowleft_{\tau=j},\omega} \cdot 0 \\
        &= \ket{i,\Vec{x}}
    \end{split}
\end{align}

\end{scenario}
Consider Equations \ref{eq:scenario-one-self-loop-target-invertion} and \ref{eq:scenario-one-self-loop-target-non-invertion}. Note that the phase of the self-loops $\ket{\circlearrowleft_{j}}$ is inverted only when $\omega = \vec{x}$, \textit{i.e.}, if $\ket{\textbf{x}}$ contains the target state. According to Equations \ref{eq:scenario-two-jth-self-loop-omega-neq-x} and \ref{eq:scenario-two-jth-self-loop-omega-eq-x}, the $j$th self-loop, where $\tau \neq j$, stay unchanged even if $\ket{\textbf{x}}$ contains the target state. The same behavior observed in Scenario \ref{sc:scenario-one} 
occurs with the $i$th non-loop edge in Scenario \ref{sc:scenario-three} according to Equations \ref{eq:scenario-three-ith-self-loop-omega-neq-x} and \ref{eq:scenario-three-ith-self-loop-omega-eq-x}. The phase will only be inverted if $\ket{\textbf{x}}$ contains the target state. As we can see, the oracle is able to partially invert the phase of the target state, based on position and edge information.

\end{document}

%% file: lib/preambule.tex
\usepackage{braket}
\usepackage{graphicx}
\usepackage{amsmath}
\usepackage{amssymb}
\usepackage{amsfonts}
\usepackage{float}
\usepackage[table,xcdraw]{xcolor}
\usepackage{cite}
\usepackage{subfig}

\usepackage{setspace}
\usepackage{multirow}